\documentclass{article}

\usepackage{PRIMEarxiv}

\usepackage[utf8]{inputenc} % allow utf-8 input
\usepackage[T1]{fontenc}    % use 8-bit T1 fonts
\usepackage{hyperref}       % hyperlinks
\usepackage{url}            % simple URL typesetting
\usepackage{booktabs}       % professional-quality tables
\usepackage{amsfonts}       % blackboard math symbols
\usepackage{nicefrac}       % compact symbols for 1/2, etc.
\usepackage{microtype}      % microtypography
\usepackage{lipsum}
\usepackage{fancyhdr}       % header
\usepackage{graphicx}       % graphics
\graphicspath{{media/}}     % organize your images and other figures under media/ folder
\usepackage{subcaption}
\usepackage{natbib}
\usepackage{amsmath}

\title{Stochastic full waveform inversion with deep generative prior for uncertainty quantification
}

\author{
  Yuke Xie, Hervé Chauris, Nicolas Desassis \\
  Center for Geosciences, Mines Paris \\
  PSL Research University \\
  77300, Fontainebleau, France\\
  \texttt{yuke.xie@minesparis.psl.eu} \\
}

\begin{document}
\maketitle

\begin{abstract}
To obtain high-resolution images of subsurface structures from seismic data, seismic imaging techniques such as Full Waveform Inversion (FWI) serve as crucial tools. However, FWI involves solving a nonlinear and often non-unique inverse problem, presenting challenges such as local minima trapping and inadequate handling of inherent uncertainties. In addressing these challenges, we propose leveraging deep generative models as the prior distribution of geophysical parameters for stochastic Bayesian inversion. This approach integrates the adjoint state gradient for efficient back-propagation from the numerical solution of partial differential equations. Additionally, we introduce explicit and implicit variational Bayesian inference methods. The explicit method computes variational distribution density using a normalizing flow-based neural network, enabling computation of the Bayesian posterior of parameters. Conversely, the implicit method employs an inference network attached to a pretrained generative model to estimate density, incorporating an entropy estimator. Furthermore, we also experimented with the Stein Variational Gradient Descent (SVGD) method as another variational inference technique, using particles. We compare these variational Bayesian inference methods with conventional Markov chain Monte Carlo (McMC) sampling. Each method is able to quantify uncertainties and to generate seismic data-conditioned realizations of subsurface geophysical parameters. This framework provides insights into subsurface structures while accounting for inherent uncertainties.
\end{abstract}

% keywords can be removed
\keywords{Inverse Problem \and Deep Learning \and Generative Models \and Full Waveform Inversion \and Bayesian Inference}

\section{Introduction}
Solving an inverse problem means to find a set of parameters that best fit observed data. Normally it starts with predicting the result of a measurement and requires a model of the system under investigation, and a physical theory linking the parameters of the model to the observations, this can typically be a partial differential equation. The prediction of observations, given the values of the parameters defining the model, constitutes the so-called forward problem. The inverse problem consists of using the results of actual observations to infer the values of the parameters characterizing the system under investigation \citep{calvetti2018inverse}.

Seismic imaging is essential for understanding the Earth subsurface, relying on recorded seismic waveforms to reveal hidden subsurface structures and parameters. It is a typical inverse problem, hence, the exploration of seismic imaging begins with the forward problem, which involves predicting seismic responses by solving the wave equation, based on known subsurface properties. 

At the basis of this effort is Full-Waveform Inversion (FWI), a modern technique that provides high-resolution images of the subsurface by exploiting information in the recorded seismic waveforms \citep{virieux2009overview, chauris2019fwi}. This approach requires solving a nonlinear and typically non-unique inverse problem. The standard FWI approach is deterministic, start from an initial guess, update the parameters through a gradient-based method. The adjoint of the partial differential equation (PDE) represents the gradient of the data residual with respect to the unknown parameters. Here, Bayesian inference serves as a helpful tool, enabling us to unravel the complexities of inverse problems such as FWI while quantifying the uncertainties associated with the results. So far, two separate approaches are classified: deterministic and stochastic. Deterministic methods aim to reveal a sole optimal solution which is referred to as the maximum a posteriori (MAP), frequently achieved through optimization methods \citep{virieux2009overview}. However, these conventional strategies can often get stuck in a local minimum and do not fully address the uncertainties inherent in seismic inversion \citep{plessix2006review}.

Bayesian inference provides a framework to solve inverse problems and quantify uncertainties. The method relies on Bayes’ theorem to start with a prior probability density function (pdf) with known information from the data to construct a so-called posterior pdf \citep{calvetti2018inverse}. While conventional sampling methods such as Markov chain Monte Carlo (McMC) \citep{metropolis1949monte, hastings1970monte} are highly effective and widely used, they can be computationally intensive for non-linear inverse problems involving the solution of physical equations and high dimensionality, such as in geophysical imaging \citep{gesret2015propagation, bottero2016stochastic, zhang2020variational}.

Deep learning has recently played a role in reshaping inversion methodologies, presenting innovative solutions to traditional challenges \citep{lecun2015deep}. Due to the improvement of computational power, we can now leverage deep learning to model and understand complex phenomena using non-linear functions. A key component of this transformation lies in the utilization of deep generative models, such as Generative Adversarial Networks (GAN) which can learn the transformation from a simple distribution to complex multivariable distributions\citep{goodfellow2020generative}, which have demonstrated considerable efficacy in diverse applications in generating realistic images such as fake face. Within the realm of generative models, Wasserstein Generative Adversarial Networks (WGAN) have emerged as a noteworthy advancement \citep{arjovsky2017wasserstein}. WGAN is inspired by optimal transport \citep{villani2009optimal, santambrogio2015optimal,peyre2019computational}, which aims to provide stable gradient by minimizing the Wasserstein distance between the generated and real probability distributions, it offers improved stability and convergence, addressing certain limitations of gradient vanishing in traditional GANs and enhancing their applicability in complex tasks \citep{arjovsky2017wasserstein, weng2019wgan}. These deep generative models, exemplified by GANs, WGANs or recently diffusion models \citep{ho2020denoising}, serve as powerful prior distributions, capable to capture the intricate nature of geophysical parameters \citep{fang2020deep, wang2023prior, bhavsar2023principled}. The integration of the numerical solution of partial differential equations with deep learning framework provides the link between the physics and the strong prior knowledge of the unknown parameters, represents a paradigmatic shift towards more probabilistic and uncertainty-aware inversion methodologies.

In this work, we parameterize the unknown subsurface parameters space by training the generative model, such that this generative model can learn the prior distribution in dimensional-reduced latent space that can generate fake subsurface images. We propose to use the variational Bayesian inference method \citep{attias1999variational} to approximate the Bayesian posterior distribution conditioned to seismic observations. It incorporates the notion of probability distributions over model parameters, capturing the inherent variability in the parameter space by approximating the target posterior with a variational distribution. More specifically, we propose to use both implicit and explicit deep learning methods for variational Bayesian inference in FWI. In the explicit method, we employ a normalizing flow \citep{rezende2015variational} based generative neural network (Figure \ref{fig:flow-network}) as a precursor to the pretrained generator $G_\theta$. The normalizing flow transforms a simple distribution into a more complex distribution, allowing for explicit variational density estimation using change of variable theorem. In contrast, the implicit method involves employing an inference network and an entropy estimator for variational density estimation, this method allows for the estimation of density without explicitly defining the probability distribution function. We solve the wave equation to obtain the predicted seismic data and compute the likelihood function, and then conduct gradient back-propagation through fully differentiable neural networks to optimize the parameters $\phi$ of the precursor. During the inference process, the pretrained generator $G_\theta$ remains fixed and is not updated. Additionally, we experimented with the Stein Variational Gradient Descent (SVGD) method \citep{liu2016stein}, which uses particles to perform variational inference. We compare all these methods with the McMC algorithm. The resulting inference methods provide a posterior distribution in the dimensional reduced latent space, enabling the generation of numerous subsurface realizations conditioned on the seismic observations. We compared the results and performance of all these four methods to evaluate their effectiveness and efficiency in capturing the uncertainties and generating accurate subsurface models. Specifically, we assessed the efficiency of each method and their capability to reconstruct multiple posterior modes.

In our exploration of this field, we find that GAN and variational Bayesian inference have potential for advancing our understanding of seismic imaging and providing a perspective on exploring Bayesian posteriors in inverse problems. Our contributions can be summarized as follows:

\renewcommand{\theenumi}{\roman{enumi}}%
\begin{enumerate}
  \item We parametrize the velocity space by training a generative model that can generate velocity fields realizations by sampling in a lower dimensional latent space.
  \item We solve a partial differential equation to obtain the seismic data residual, and solve the adjoint state equation to provide back-propagation of the gradient through the fully differentiable generative neural network.
  \item We introduce both explicit and implicit deep learning methods for variational Bayesian inference, as well as the particle method SVGD. We carry out synthetic experiments of stochastic FWI using a deep generative model as prior, and compare these variational Bayesian inference methods with the McMC sampling method.
  \item We obtained the Bayesian posterior distribution of velocity fields conditioned to seismic observations.
\end{enumerate}

\section{Theory}
\label{sec:theory}
\subsection{Bayesian Inference}
\label{sec:bayesian_inference}
Bayesian inference is fundamental in addressing inverse problems, a prevalent challenge in scientific fields such as seismic imaging. At its essence, Bayesian inference provides a systematic methodology to enhance our knowledge of the unknown latent variables $\mathbf{z}$ using the observed seismic data $d_{\text{obs}}$. The posterior probability distribution is expressed as

\begin{equation}
\label{eq:bayes}
P(\mathbf{z}|d_{\text{obs}}) \propto P(d_{\text{obs}}|\mathbf{z}) \cdot P(\mathbf{z}),
\vspace{+6pt}
\end{equation}
where $P(\mathbf{z}|d_{\text{obs}})$ is the posterior distribution of the parameter given the observed data $d_{\text{obs}}$, $P(d_{\text{obs}}|\mathbf{z})$ represents the likelihood of the observed data given the parameter, and $P(\mathbf{z})$ denotes the prior distribution conveying our initial beliefs regarding the parameter. The proportionality underscores that the posterior is directly proportional to the product of the likelihood and the prior. The likelihood of the observed data given the parameter can be written as

\begin{equation}
\label{eq:likelihood}
P(d_{\text{obs}}|\mathbf{z}) \propto \exp\left(\frac{-\left\| d_{\text{pred}} - d_{\text{obs}} \right\|^2}{2\sigma^2}\right) ,
% = \exp\left(-\frac{J(z)}{\sigma^2}\right),
\vspace{+6pt}
\end{equation}
where $d_{\text{pred}}$ represents the predicted data based on the physical model, $d_{\text{obs}}$ signifies the observed data, and $\sigma$ denotes the noise level inherent in the observation. The likelihood function quantifies the probability of observing the data given a particular model parameterization $\mathbf{z}$. It measures the agreement between the predicted and observed data, with higher probabilities indicating better agreement. 
% The term $J(z)$ quantifies the residual between the observed and modeled data, serving as a measure of the discrepancy between the two.

The prior distribution $P(\mathbf{z})$ encapsulates our prior beliefs about the parameter $\mathbf{z}$ and plays a crucial role in Bayesian inference. In the context of Gaussian distributions, the prior distribution can be expressed as

\begin{equation}
\label{eq:prior}
P(\mathbf{z}) \propto \exp\left(-\frac{1}{2}\mathbf{z}^T\frac{1}{\sigma^2}\mathbf{z}\right) = \exp \left(-\frac{\left\|\mathbf{z}\right\|^2}{2}\right),
\vspace{+6pt}
\end{equation}
where $\sigma$ represents the standard deviation of the prior distribution. This term incorporates our knowledge or assumptions about the distribution of the parameter $\mathbf{z}$ before observing any data. It serves as a regularization term, influencing the posterior distribution by favoring solutions that are consistent with our prior beliefs.

Inverse problems like in seismic imaging, often face the challenge of non-uniqueness. This means that there can be multiple parameter configurations that adequately explain observed data. Bayesian inference provides the theory to address this challenge by providing a probabilistic characterization of inferred parameters. Instead of giving a single deterministic solution, Bayesian inference offers a distribution over potential parameter values. This helps in quantifying uncertainties inherent in the inferred model. Essentially, Bayesian inference provides a systematic method for solving inverse problems in geophysical seismic imaging. By combining prior knowledge with observed data, Bayesian inference not only helps estimate model parameters but also offers a framework for addressing uncertainties. This is essential for decision making in various scientific and engineering fields.

\subsection{Adjoint State Method}
In the context of seismic imaging, FWI is a common approach to reconstruct subsurface models by comparing observed seismic data with simulated data generated from a forward model \citep{virieux2009overview, chauris2019fwi}. This process involves solving the wave equation, which describes the propagation of seismic waves through the subsurface parameters such as velocity. The 2D acoustic wave equation of P-wave velocity used in this study can be written as
\begin{equation}
\frac{1}{m^2(x)}\frac{\partial^2 p}{\partial t^2} - \Delta p = S(t) \delta(s - x),
\vspace{+6pt}
\label{eq:acoustic}
\end{equation}
where $x$ represents 2D spatial positions, \(t\) represents time, \(s\) represents the source position usually at the surface, \(p(s, x, t)\) is the pressure wave field, \(S(t)\) is the source wavelet, \(\delta()\) is the Dirac distribution indicating the source spatial position, and \(m(x)\) is the velocity field indicating wave propagation velocity at specific positions. Solving this equation provides the time-dependent pressure field in the subsurface due to seismic sources, and the amplitude at specific receivers location \citep{virieux2009overview}.

The predicted shot records are obtained by solving an individual wave equation per shot location and depend on the parametrization \( m(x) \) of our wave propagator. In standard FWI, the most common function for measuring the data residual between the observed and modeled data in Bayesian inversion is the \(\ell^2\) norm, which leads to the following objective function

\begin{equation}
\underset{m}{\text{minimize }} J(m) = \frac{1}{2} \sum_{i=1}^{n_s} \left\| d_{\text{pred}}^{i}(m, x^{i}) - d_{\text{obs}}^{i} \right\|_2^2,
\vspace{+6pt}
\end{equation}
where the index \( i \) runs over the total number of shots \( n_s \). The gradient-based optimization entails computing the gradient of the objective function \( J \) with respect to the model parameters \( m \):
\begin{equation}
\nabla J = \frac{\partial J}{\partial m}.
\end{equation}

FWI employs the adjoint state method to efficiently compute the gradient of the objective function, which quantifies the residual between observed and simulated data \citep{plessix2006review}. The essence of the method lies in its ability to compute the gradient by solving both the forward and adjoint wave equations. Extending the objective function by forming up the Lagrangian leads to

\begin{equation}
\begin{split}
J_{ext}[m,p&,\lambda] = \frac{1}{2}\left\| d_{\text{pred}}(m, x) - d_{\text{obs}} \right\|_2^2 \\
& -  \left< \lambda,  \left( \frac{1}{m^2(x)}\frac{\partial^2 p}{\partial t^2}-\Delta p \right)-S(t)\delta(s-x)\right>_{s,x,t}
\end{split}
\label{eq:adjoint}
\end{equation}

By differentiating $J_{ext}[m,p,\lambda]$ with respect to $p$ leads to the corresponding adjoint wave equation, 

\begin{equation}
\label{eq:adjoint_equation}
\frac{1}{m^2(x)}\frac{\partial^2 \lambda}{\partial t^2} - \Delta \lambda = <M, Mp-d_\text{obs}>
\end{equation}
which is the transposed version of the forward wave equation, the fact that in the process of deriving the adjoint equation from the data residual function, the forward equation is essentially reversed, which is the similar to the back-propagation process for derivation in neural networks \citep{lecun2015deep}. The data residual on the right side of adjoint equation serves as the sources term in the forward wave equation, and is back-propagated thanks to the transposed equation. By differentiating extended Lagrangian with respect to $m$ leads to the gradient 
\begin{equation}
\nabla J = \frac{\partial J}{\partial m} = \frac{2}{m^3} \iint ds \, dt \, \lambda(s, x, t) \frac{\partial^2 p}{\partial t^2}(s, x, t),
\end{equation}
for discrete time steps and shots, the gradient is computed as

\begin{equation}
\frac{\partial J}{\partial m} = \frac{2}{m^3} \sum_{i=1}^{n_s} \sum_{t=1}^{n_t} \lambda(s, x, t) \frac{\partial^2 p}{\partial t^2}(s, x, t),
\vspace{+6pt}
\end{equation}
where \( \lambda(s, x, t) \) is the time-reversed adjoint wavefield, $\frac{\partial^2 p}{\partial t^2}$ is the second temporal derivative of the forward wavefield which can be estimated with a second-order finite-difference approach, thereby corresponds to performing the point-wise multiplication of the adjoint wavefield with the second time derivative of the forward wavefield and summing over all time
steps. $\nabla = \frac{\partial J}{\partial m} \in \mathbb{R}^{m}$ is the gradient which we back-propagate to neural network parameters using chain rule. The numerical solution of the PDE is performed with GPU-accelerated computing using CuPy \citep{cupy_learningsys2017}.

\section{Generative Adversarial Network}
\label{sec:generative}

\begin{figure}
    \centering
    \includegraphics[width=0.75\linewidth]{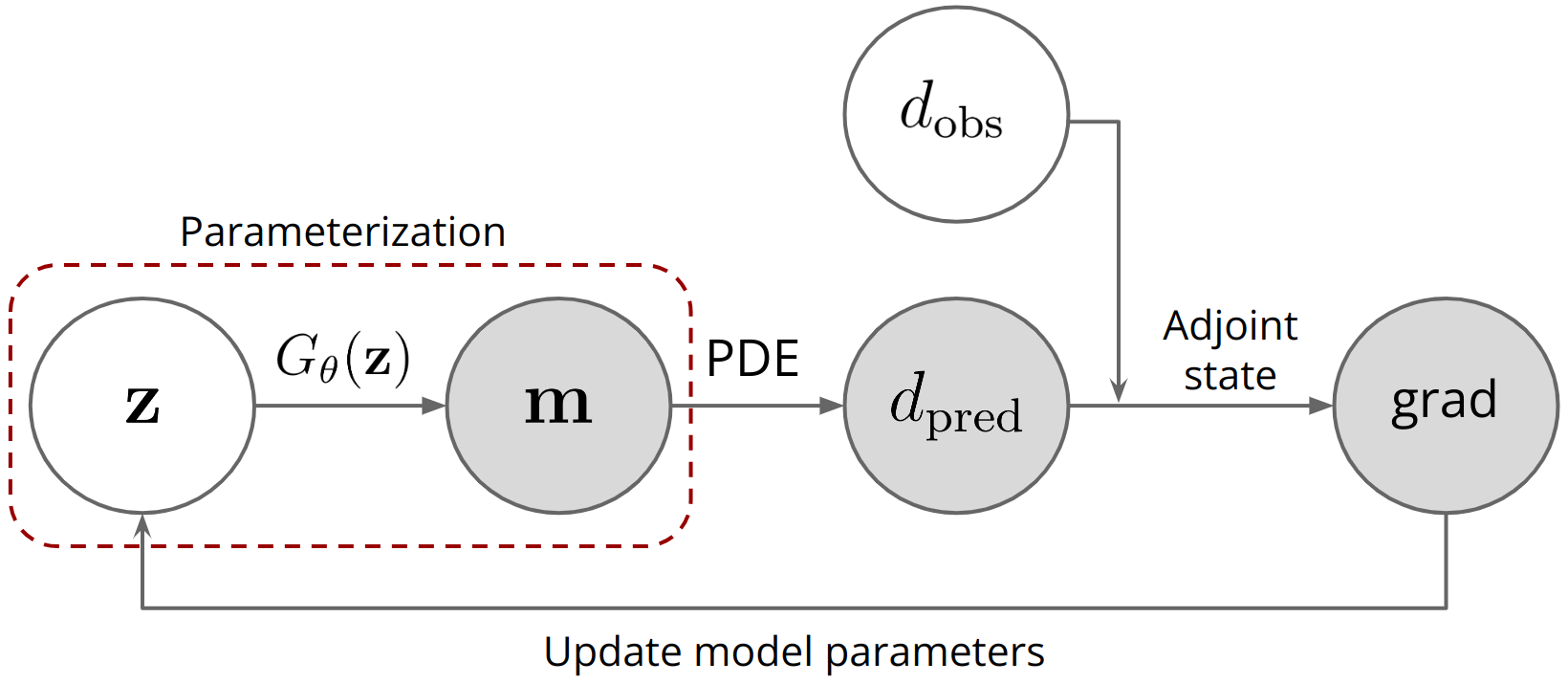}
    \caption{Diagram illustrating Full Waveform Inversion with Parameterization, we parameterize the subsurface velocity field $\textbf{m} = G_\theta(\mathbf{z})$, and we solve acoustic wave equation to predict the synthetic seismic data $d_\text{pred}$. By evaluating the data residual with observed seismic data $d_\text{obs}$, we obtain the gradient using adjoint state method and subsequently update model parameters using optimization algorithms.}
    \label{fig:parameterization}
\end{figure}

Parameterization is useful in geophysical modeling. In this study, we employ a deep learning model to parameterize the subsurface model, enabling the generation of stochastic realizations of subsurface images. The probabilistic distribution of these subsurface images is represented by transforming a dimensional-reduced multivariate Gaussian distribution using deep learning. More specifically, we train a generative model GAN to produce model parameters $m = G_{\theta}(\mathbf{z})$ (Figure \ref{fig:parameterization}) by sampling latent variable vectors $\mathbf{z}$. These representations Bayesian prior knowledge of subsurface structures. GANs generate data distributions based on examples, requiring a generator $G_{\theta}(\mathbf{z})$ and a discriminator $D_{\omega}(\mathbf{m})$. The generator produces samples mimicking the training data, while the discriminator distinguishes real from generated samples \citep{goodfellow2020generative}. Both are trained via minimize and maximize

\begin{equation}
\begin{aligned}
&\min_{\theta} \max_{\omega} L(D_{\omega}, G_{\theta}) \\
& = \mathbb{E}_{\mathbf{m} \sim p_{r}(\mathbf{m})} [\log D_{\omega}(\mathbf{m})] + \mathbb{E}_{\mathbf{z} \sim p_\mathbf{z}(\mathbf{z})} [\log(1 - D_{\omega}(G_{\theta}(\mathbf{z})))] \\
& = \mathbb{E}_{\mathbf{m} \sim p_{r}(\mathbf{m})} [\log D_{\omega}(\mathbf{m})] + \mathbb{E}_{\tilde{\mathbf{m}} \sim p_g(\tilde{\mathbf{m}})} [\log(1 - D_{\omega}(\tilde{\mathbf{m}})]
\end{aligned}
\end{equation}
where $p_{\textbf{z}}$ is data distribution over noise input $\textbf{z}$, $p_g$ is the generator distribution over generated image $\tilde{\mathbf{m}}$, $p_r$ is the distribution over real data sample $r$.
Training GANs is inherently unstable, and finding stable training methods remains an open research problem \citep{weng2019wgan}. In this work, we employ a so-called Wasserstein-GAN (WGAN) \citep{arjovsky2017wasserstein, weng2019wgan}, inspired by optimal transport, which aims to provide stable gradient by minimizing the Wasserstein distance \citep{peyre2019computational} between the generated and real probability distributions. However, standard WGAN still suffers from unstable training like slow convergence and vanishing gradients after a so-called weight clipping method as a way to enforce a Lipschitz constraint \citep{weng2019wgan}. \cite{gulrajani2017improved} propose a new method WGAN with Gradient Penalty (WGAN-GP) to replace weight clipping method with a gradient penalty term to enforce a Lipschitz constraint. The new objective for discriminator is:

\begin{equation}
\begin{split}
L =\ & \mathbb{E}_{\tilde{\mathbf{m}} \sim p_g(\tilde{\mathbf{m}})}[D_{\omega}(\tilde{\mathbf{m}})] - \mathbb{E}_{\mathbf{m} \sim p_{r}(\mathbf{m})} [D_{\omega}(\mathbf{m})] \\&+ \lambda \ \mathbb{E}_{\tilde{\mathbf{m}} \sim p_g(\tilde{\mathbf{m}})}\left[(||\nabla_{\hat{\mathbf{m}}} D_{\omega}(\hat{\mathbf{m}})||_2 - 1)^2\right]
\end{split}
\vspace{+10pt}
\end{equation}
% $\hat{m} \sim \mathbf{P}_{\hat{m}}$
where $\lambda$ is a weighting factor on the gradient penalty and we choose $\lambda = 10$ in this study. And $\hat{\mathbf{m}} = \mathbf{m} + (1-\epsilon)\tilde{\mathbf{m}}$ is a random sample between a real and generated sample controlled by a random variable $\epsilon \sim \text{U}(0,1)$.

WGAN-GP lies in its unique architecture wherein a gradient penalty is incorporated into the loss function. This addition introduces a regularization mechanism during training, aiming to mitigate issues encountered in traditional GAN training, such as mode collapse and instability. WGAN-GP improves the stability and convergence of the GAN training process. The gradient penalty acts as a regularizer, encouraging smoother learning dynamics and preventing the discriminator from becoming overly dominant, thereby promoting a more balanced adversarial training scenario.

\begin{figure*}
     \centering
     \begin{subfigure}[b]{0.495\textwidth}
         \centering
         \includegraphics[width=\textwidth]{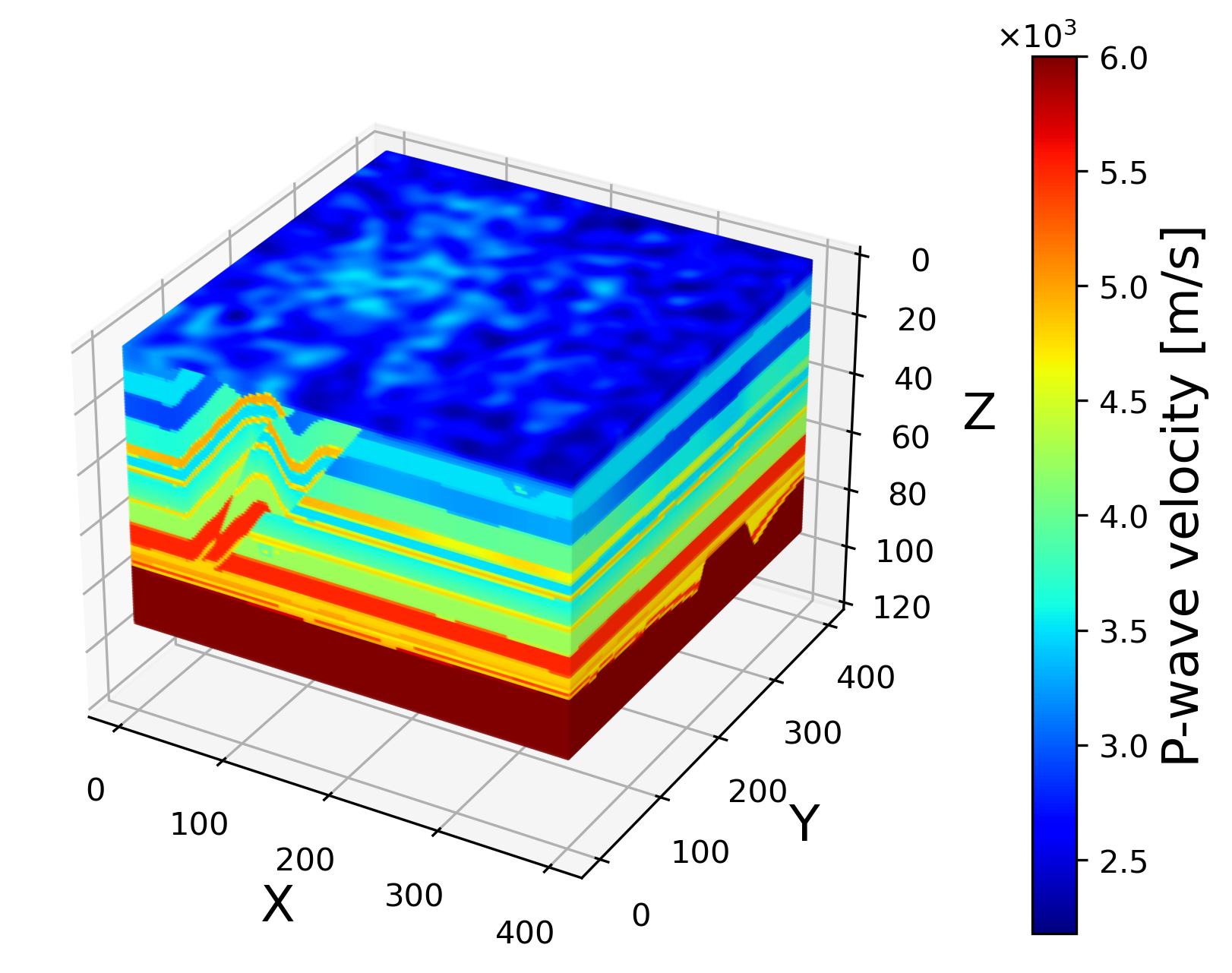}
         \caption{3D Overthrust model}
         \label{fig:Overthrust}
     \end{subfigure}
     \hfill
     \begin{subfigure}[b]{0.495\textwidth}
         \centering
         \includegraphics[width=\textwidth]{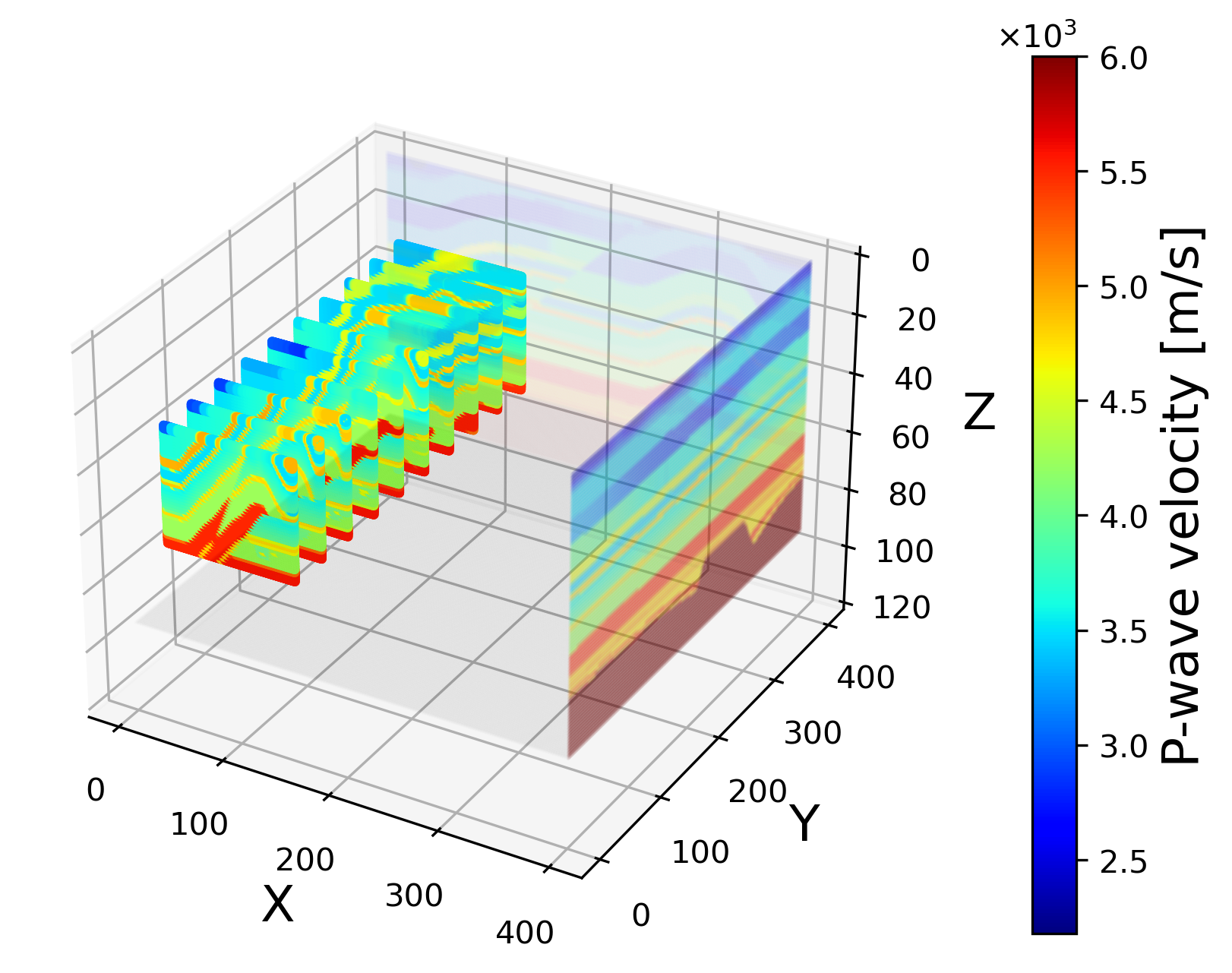}
         \caption{Training data}
         \label{fig:Training_data}
     \end{subfigure}
     
     \begin{subfigure}[b]{1\textwidth}
         \centering
         \includegraphics[width=\textwidth]{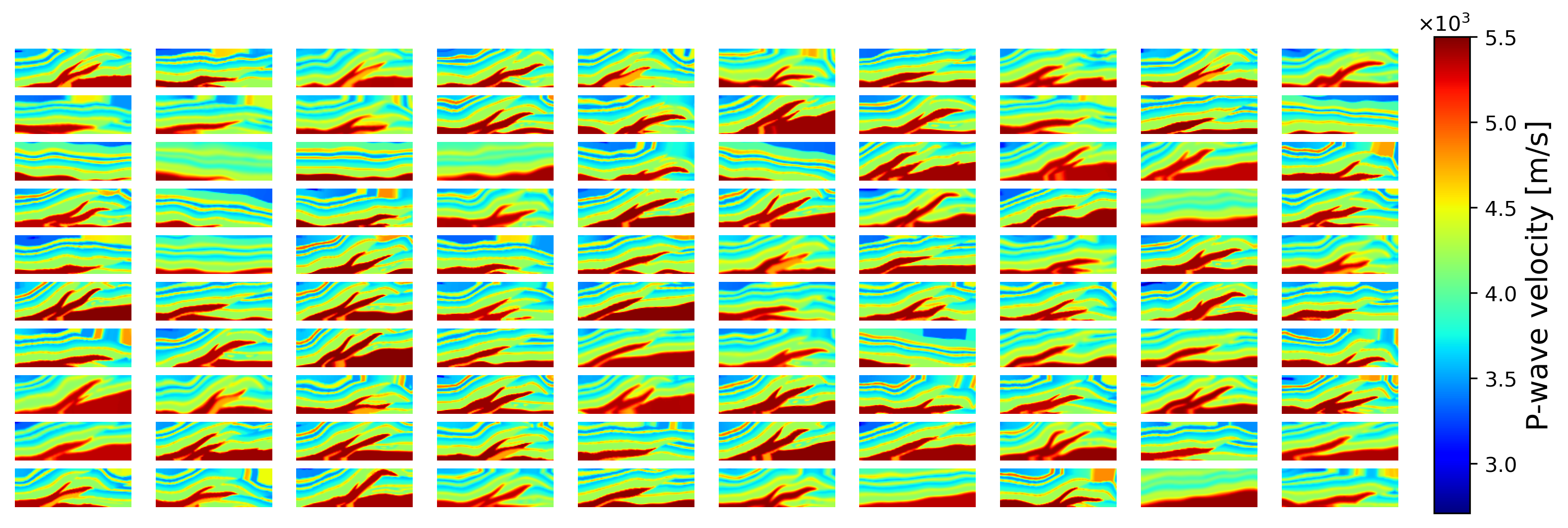}
         \caption{Image-data-generator (training) images}
         \label{fig:Training}
     \end{subfigure}
     
     \begin{subfigure}[b]{1\textwidth}
         \centering
         \includegraphics[width=\textwidth]{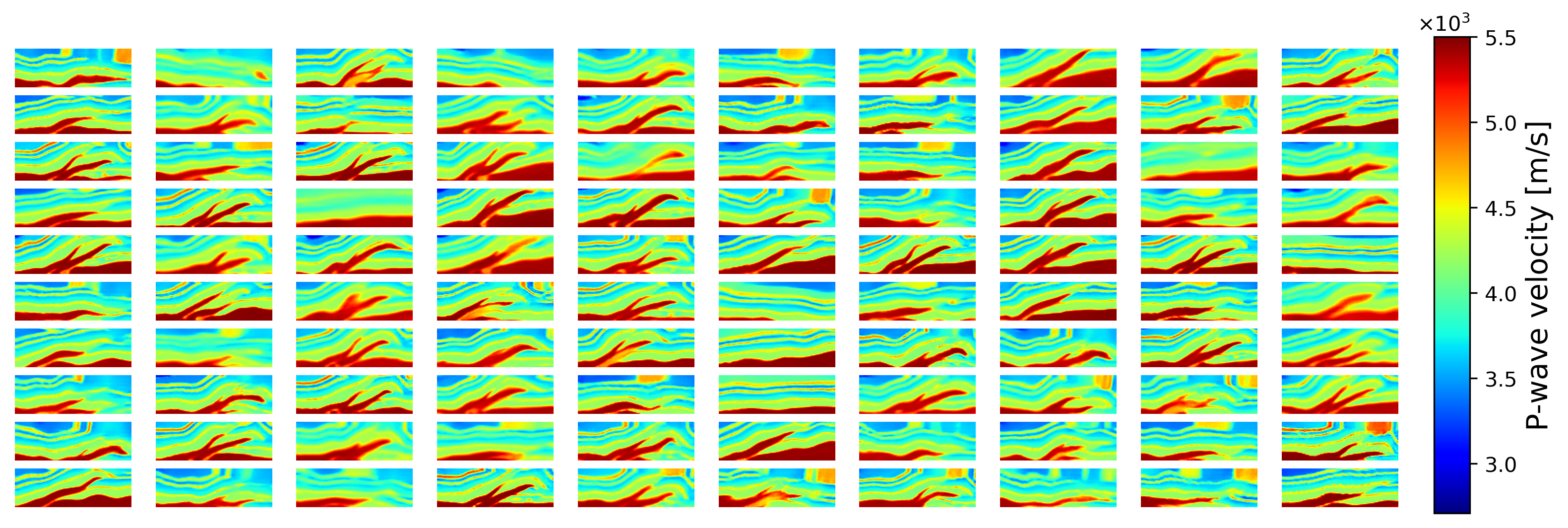}
         \caption{GAN generated images}
         \label{fig:dataset_and_generated}
     \end{subfigure}
     
     \caption{Overview of the dataset and generated images in this work. \textbf{a} 3D Overthrust dataset with dimensions $94\times401\times401$ and \textbf{b} training images cropped from the dataset, with dimensions $40\times120$. \textbf{c} Training images and \textbf{d} generated fake images from GAN.}
     \label{fig:gan}
\end{figure*}

In this study, we employed the 3D Overthrust dataset (Figure \ref{fig:gan}a) for training our GAN. This dataset is widely acknowledged in the seismic imaging domain, it represents the intricate geological formations in the subsurface. To ensure uniformity and focus on specific subsurface features, we cropped the dataset using a sliding window at different cross-sections along y-axis and result in the extraction of 401 original images cropped from the dataset (Figure \ref{fig:gan}b), each measuring 40-depth x 120-width, capturing the synthetic subsurface characteristics including a geological interface for our research objectives. To obtain more images and more variability for the training of the GAN, we created an image-data-generator using data augmentation methods. To be specific, during each training iteration, we augment the training dataset by generating random variations of the original images cropped from the overthrust dataset. These variations include rotations within the range of -10° to +10°, horizontal shifts by -2 to +2 pixels, shearing within the range of -2 to +2 radians, elastic transformations for distortion, and nearest interpolation to fill edges. The image-data-generator includes a variety of geological layers (Figure \ref{fig:gan}c), displaying P-wave velocity ranging from 2.1 km/s to 5.5 km/s. These images represent different geological structures for detailed analysis and modeling.

The key parameters employed for training the WGAN-GP included utilizing an Adam optimizer with a learning rate of 1e-5, $\beta_1 = 0.5$, and $\beta_2 = 0.9$. The model underwent extensive training for 50,000 epochs with a batch size of 256 for each iteration. 
The WGAN was configured with a latent space dimensionality of 100. Training consisted of iterative updates alternating between the generator and discriminator networks. Notably, the discriminator underwent five additional training iterations compared to each generator update to enhance its discriminatory capabilities. Both networks were designed with convolutional kernels to capture spatial features and patterns in the dataset.

\begin{figure*}
    \begin{subfigure}[b]{1\textwidth}
    \centering
    \includegraphics[width=0.4\textwidth]{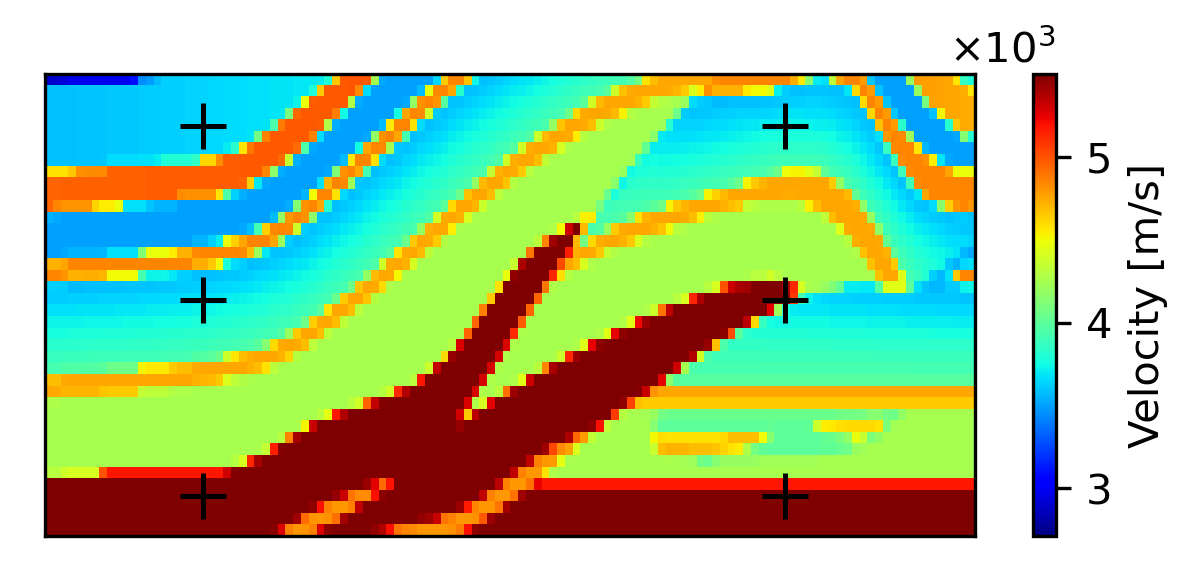}
    % \caption{Selected locations to compute correlation.}
    \label{fig:correlation_locations}
    \end{subfigure}
    \begin{subfigure}[b]{1\textwidth}
    \centering
    \includegraphics[width=0.7\textwidth]{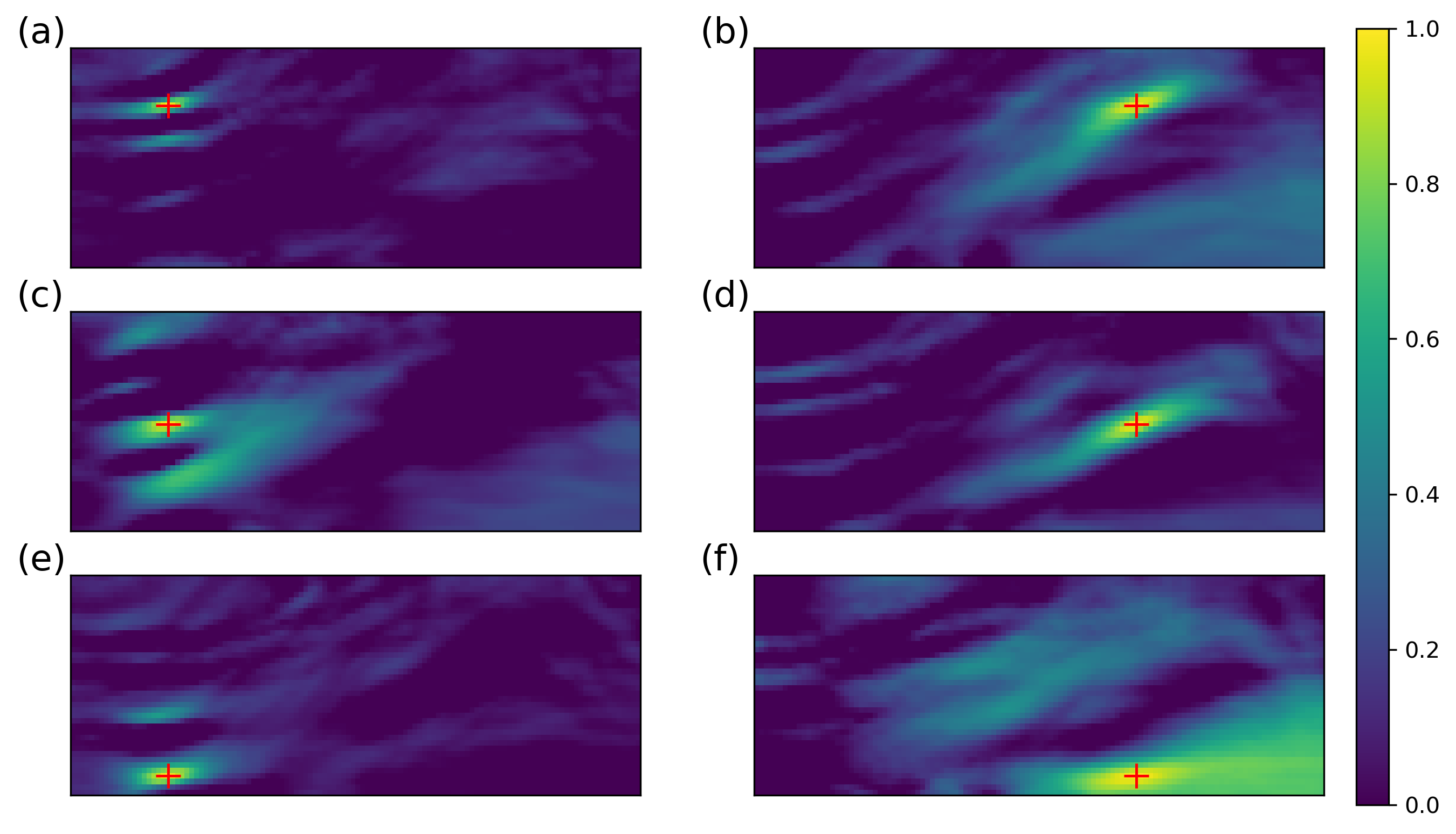}
    \end{subfigure}
    \caption{Correlation between grids of the GAN-generated images. The cross indicates the grid point used to compute the correlation.}
    \label{fig:correlation_prior}
\end{figure*}

% \begin{figure*}
% \centering
% \includegraphics[width=0.7\textwidth]{images/prior/PCA_prior.png}
% \caption{Visualization of training images and GAN-generated images in 2D using principal component analysis (PCA). This visualization is achieved by applying PCA to the two principal components of the GAN-generated image space. The plot illustrates the distribution of images in a reduced feature space}
% \label{fig:pca}
% \end{figure*}

The resulting pretrained generator successfully learned to generate images representative of the seismic velocity dataset. Figure \ref{fig:gan}c, d showcases both the original training images and the images generated by the pretrained GAN, demonstrating the model learned features. The learned patterns and distributions from this dataset operate as prior knowledge in subsequent subsurface inversion processes. 

The correlation analysis presented in Figure \ref{fig:correlation_prior} illustrates the consistency and coherence between various grids of the GAN-generated images. The cross symbol indicates the specific grid location used to compute the correlation. Figures \ref{fig:correlation_prior}b and d indicate the presence of slant structures within the Overthrust dataset, exhibiting notable correlations in the slant direction. Figures \ref{fig:correlation_prior}a, c, and e showcase significant correlations observed across different layers in depth, indicative of the GAN capacity to capture this layered information effectively. Figure \ref{fig:correlation_prior}f highlights a distinct pattern in the lower part of the image space characterized by elevated correlations in the horizontal direction, thereby suggesting the presence of continuous high-velocity layers in deeper regions of the model. The correlations ensure that the GAN can capture the underlying geological features present in the training dataset. 

By leveraging this prior knowledge, it aids in deducing accurate subsurface structural parameters from observed seismic data, providing valuable prior knowledge into the subsurface parameters.

\section{Reconstructing posterior}
\label{inference}
Bayesian inference involves constructing a posterior pdf \( p(\mathbf{z}|d_{\text{obs}}) \) for model parameters \( \mathbf{z} \) given observed data \( d_{\text{obs}} \). In Bayesian inference, Markov chain Monte Carlo (McMC) algorithms are commonly employed to sample the posterior distribution. These algorithms generate a sequence (chain) of successive samples from the posterior pdf by traversing parameter space in a structured random walk. The Metropolis–Hastings algorithm \citep{metropolis1949monte, hastings1970monte} is a well-known example. However, the basic Metropolis–Hastings algorithm may encounter computational challenges in high-dimensional spaces if the chain becomes trapped in local minima, hindering exploration of all possible minima. To address this issue, advanced gradient-based McMC methods such as Hamiltonian Monte Carlo \citep{duane1987hybrid, betancourt2017conceptual} and Langevin Monte Carlo methods \citep{roberts1996exponential, mosser2020stochastic} have been applied in geophysics.

To generate samples from the posterior distribution $\log p(\mathbf{z}|d_{\text{obs}})$ using our pretrained GAN, we employ the Metropolis-Adjusted Langevin Algorithm (MALA) with multiple chains to propose a new state $\mathbf{z}^*$ from the current state $\mathbf{z}_{t}$.

\begin{equation}
\mathbf{z}^* = \mathbf{z}_{t} + \frac{\epsilon_{t}^2}{2} \nabla \log p(\mathbf{z}_{t}|d_{\text{obs}}) + \epsilon_{t} \eta_{t},
\vspace{+6pt}
\end{equation}
where $\eta_{t} \sim \mathcal{N} (0, I)$ is a sample from a Gaussian distribution to provide randomness at each epoch, resulting in a mean $\mu(\mathbf{z}^*) = \mathbf{z}_{t} + \frac{\epsilon_{t}^2}{2} \nabla \log p(\mathbf{z}_{t}|d_{\text{obs}})$. $\epsilon_{t}$ is the step size at iteration $t$. In the context of seismic imaging, we can reformulate the equation

\begin{equation}
\mathbf{z}^* = \mathbf{z}_{t} + \frac{\epsilon_{t}^2}{2} \nabla \left( \frac{\left\| d_{\text{pred}}(G(\mathbf{z})) - d_{\text{obs}} \right\|^2}{2\sigma^2}  +
\frac{\left\|\mathbf{z}\right\|^2}{2}\right) + \epsilon_{t} \eta_{t}.
\end{equation}

A Metropolis-Hastings acceptance probability is calculated after a step,

\begin{equation}
\alpha = \frac{\pi(\mathbf{z}^*) q(\mathbf{z}_t | \mathbf{z}^*)}{\pi(\mathbf{z}_t) q(\mathbf{z}^* | \mathbf{z}_t)},
\vspace{+6pt}
\end{equation}
where $q(\mathbf{z}_2 | \mathbf{z}_1) = \mathcal{N}(\mathbf{z}_2 | \mu(\mathbf{z}_1), \epsilon I)$ and $\mathbf{z}^*$ is accepted with probability $\min(1, \alpha)$.

When employing the MALA algorithm or most McMC methods, each newly proposed state $\mathbf{z}^*$ necessitates the computation of the gradient of the seismic data residual with respect to $G(\mathbf{z})$. Though the gradient can generally be computed using the adjoint state method, it remains computationally intensive. Subsequently, both the gradient of the data residual and the gradient of the prior term $\frac{\left|\mathbf{z}\right|^2}{2}$ are propagated backward through the neural network using the automatic differentiation.

\subsection{Variational Bayesian Inference}
% However, conventional sampling methods such as the McMC algorithms are not efficient for high dimensionality. 
Variational Bayesian inference is a Bayesian inference method using optimization, it seeks an optimal approximation \( q_{\phi}^* (\mathbf{z}) \) by minimizing the Kullback–Leibler (KL)  divergence \citep{kullback1951information} between \( q_{\phi}(\mathbf{z}) \) to the posterior pdf \( p(\mathbf{z}|d_{\text{obs}}) \) within a pre-defined family of known probability distributions \( Q = \{q_{\phi}(\mathbf{z}), \phi \in \mathbb{R}^{d} \} \), where $d$ is the latent dimension 

\begin{equation}
q_{\phi}^* (\mathbf{z}) = \arg\min_{q_{\phi} \in Q} \text{KL}\left[q_{\phi}(\mathbf{z}) \, \big\| \, p(\mathbf{z}|d_{\text{obs}}) \right].
\end{equation}

\begin{figure}
    \centering
    \includegraphics[width=0.75\linewidth]{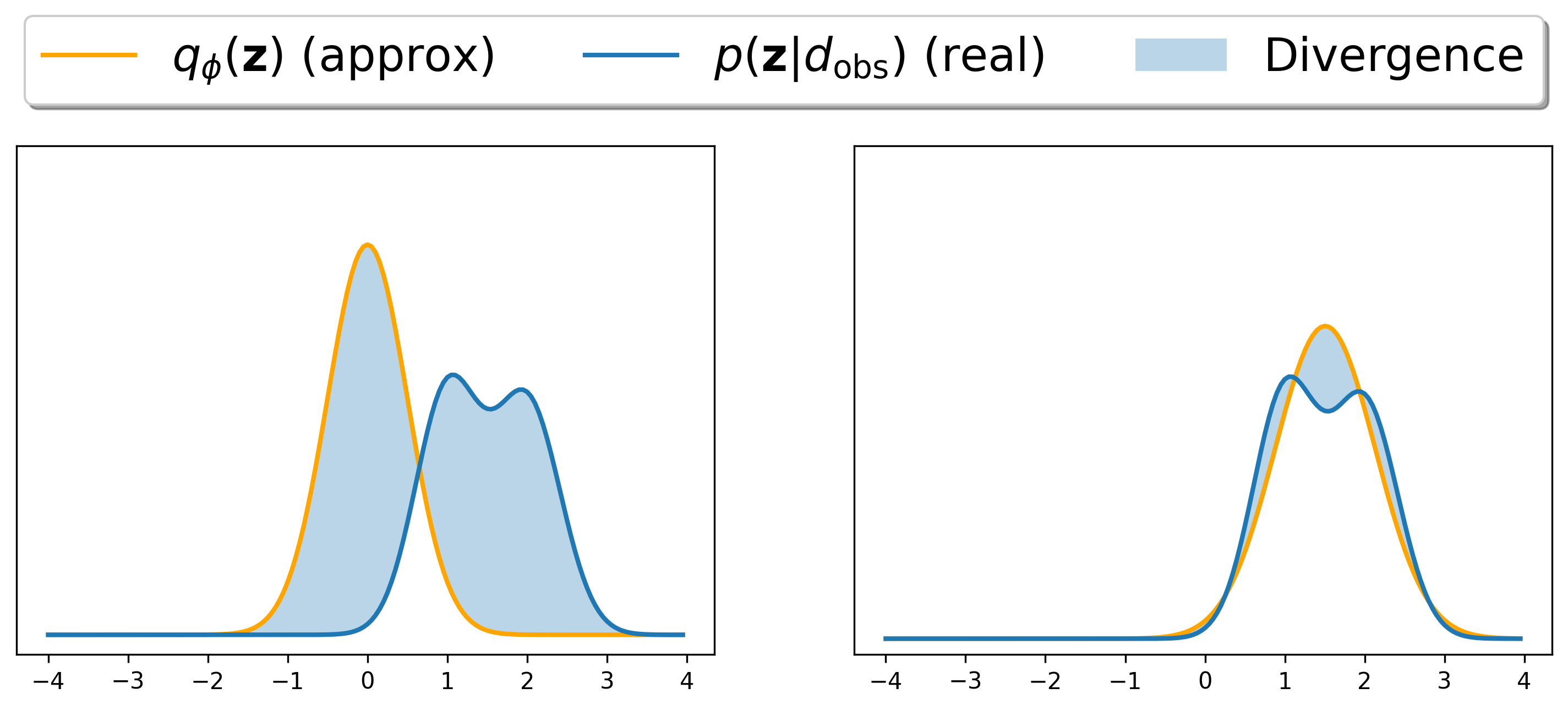}
    \caption{Minimizing KL Divergence $\text{KL}\left[q_{\phi}(\mathbf{z}) \, \big\| \, p(\mathbf{z}|d_{\text{obs}})\right]$ using gradient descent with the TensorFlow Adam optimizer to approximate a Gaussian mixture in 1D space. This is an illustration explaining the process of minimizing KL Divergence in variational Bayesian inference. The left figure shows distributions at the first epoch, and the right figure shows the approximation result.}
\end{figure}

The KL divergence measures the difference between two probability distributions and can be expressed as

\begin{equation}\label{eq:KL-1}
\begin{split}
\text{KL}\left[q_{\phi}(\mathbf{z}) \, \big\| \, p(\mathbf{z}|d_{\text{obs}})\right] &= \mathbb{E}_{\mathbf{z} \sim q_{\phi}} \left [\log q_{\phi}(\mathbf{z})\right ] \\&- \mathbb{E}_{\mathbf{z} \sim q_{\phi}} [\log p(\mathbf{z}|d_{\text{obs}})],
\end{split}
\vspace{+6pt}
\end{equation}
where the expectation is taken with respect to the distribution \( q_{\phi}(\mathbf{z}) \). The KL divergence is non-negative and only equals zero when \( q_{\phi}(\mathbf{z}) = p(\mathbf{z}|d_{\text{obs}}) \) \citep{kullback1951information}. Expanding the posterior pdf by combining equation \ref{eq:bayes} and Equation \ref{eq:KL-1}, the KL divergence becomes

\begin{equation}
\begin{split}
    \text{KL}\left[q_{\phi}(\mathbf{z}) \, \big\| \, p(\mathbf{z}|d_{\text{obs}})\right] &= \mathbb{E}_{\mathbf{z} \sim q_{\phi}} \left [\log q_{\phi}(\mathbf{z})\right ] \\&- \mathbb{E}_{\mathbf{z} \sim q_{\phi}} \left [\log p(d_{\text{obs}}|\mathbf{z})\right ] \\&- \mathbb{E}_{\mathbf{z} \sim q_{\phi}} \left [\log p(\mathbf{z}) \right] + \textbf{c},
\end{split}
\label{eq:KL-2}
\end{equation}

In the context of seismic imaging, the objective function can be constructed by combining Equation \ref{eq:KL-2} with Equations \ref{eq:likelihood} and \ref{eq:prior}. This yields

\begin{equation}
\begin{split}
\text{KL}\left[q_{\phi}(\mathbf{z}) \, \big\| \, p(\mathbf{z}|d_{\text{obs}})\right] &= \mathbb{E}_{\mathbf{z} \sim q_{\phi}} \left [\log q_{\phi}(\mathbf{z})\right ] \\&+
\mathbb{E}_{\mathbf{z} \sim q_{\phi}} \left [\frac{\left\| d_{\text{pred}}(G(\mathbf{z})) - d_{\text{obs}} \right\|^2}{2\sigma^2}\right] \\&+
\mathbb{E}_{\mathbf{z} \sim q_{\phi}} \left [\frac{\left\|\mathbf{z}\right\|^2}{2} \right] + 
\textbf{c},
\end{split}
\label{eq:kl-objective}
\vspace{+6pt}
\end{equation}
where we choose initial variational distribution $q_{\mathbf{z}}$ as a multivariate Gaussian. The noise term $\sigma$ modulates the influence of the deep generative prior on the sampling process of Bayesian inference. For the seismic data residual $\left\| d_{\text{pred}}(G(\mathbf{z})) - d_{\text{obs}} \right\|^2$, we use 2D acoustic wave equation for velocity P-wave propagation in our experiment, and solve adjoint wave equation (Equation \ref{eq:adjoint_equation}) \citep{plessix2006review} to obtain a similar neural networks back-propagation equation \citep{lecun2015deep} at the final layer of the fixed pretrained generator, providing the gradient of the data residual back to the fully differentiable neural networks. The third term $\frac{\left\|\mathbf{z}\right\|^2}{2}$ in Equation \ref{eq:kl-objective} is the regularization term from the multivariate Gaussian prior which penalize the sampling far to the mean.

One of the focus of this study is to estimate the density of the variational distribution $\log q_{\phi}(\mathbf{z})$. However, a key challenge lies in the fact that the pdf of the variational distribution is unknown and can be complex. In the following sections, we introduce both different variational methods to compute the pdf of the variational distribution.

\subsection{Inference Network}
\begin{figure*}
    \centering
    \includegraphics[width=1\linewidth]{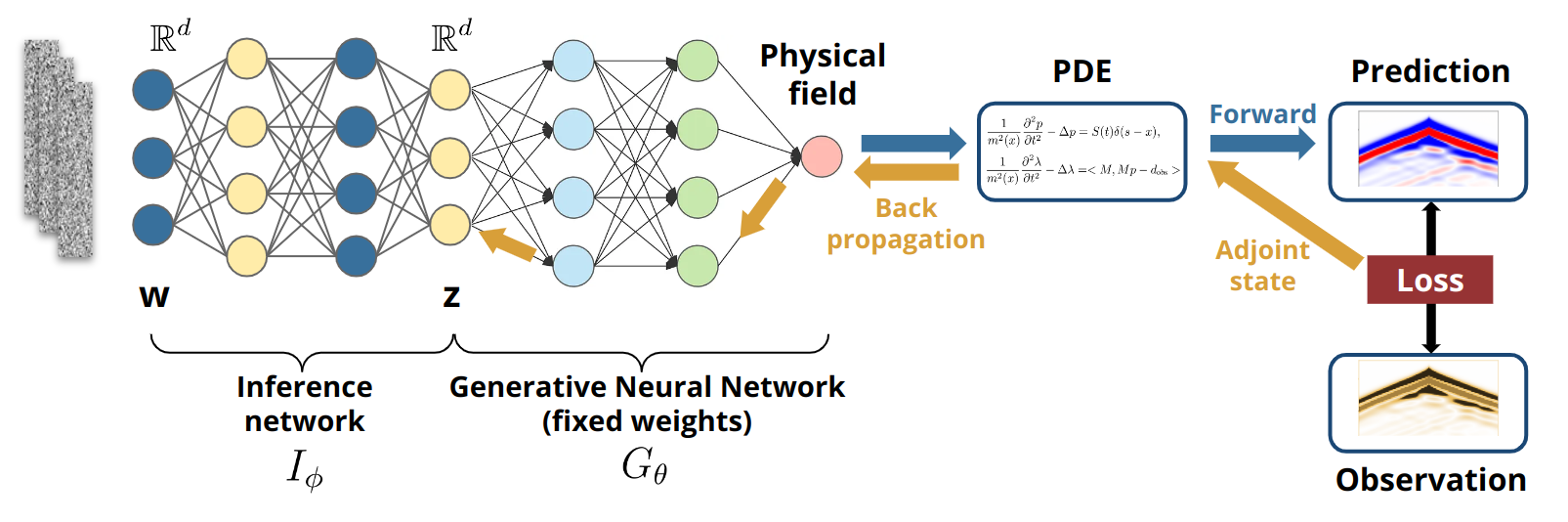}
    \caption{Structure of the implicit inference method using inference network}
    \label{fig:inference-network}
\end{figure*}

In \cite{chan2019parametric}, the authors utilize an inference neural network $I_\phi$ such that the family of parametric distribution $Q$ is described by the set of all the functions defined by the neural network $I_\phi$ applied to a given set of random variables. The parametric distribution $q_\phi$ is defined by a fully connected neural network that transforms the multi-dimensional Gaussian vector $\textbf{w}$ into another vector $\mathbf{z}$, which will be the input of the pretrained generative model (Figure \ref{fig:inference-network}). Both the input latent space and the output latent space share the same dimensionality. 

The neural network $I_\phi$, parametrized by the parameters $\phi$, embodies the trainable weights within its architecture. These parameters $\phi$ must undergo training to attain optimality. The optimal weights $\phi^*$ are those that minimize the Kullback-Leibler divergence between the variational distribution and the true posterior distribution.

% For the second term in Equation \ref{eq:kl-objective} seismic data residual $\left\| d_{\text{pred}}(G(\mathbf{z})) - d_{\text{obs}} \right\|^2$, we use 2D acoustic wave equation for velocity P-wave propagation in our experiment, and solve adjoint wave equation \ref{eq:adjoint_equation} at the final layer of the pretrained generator, providing the gradient of the data residual back through the fully differentiable generator $G$ to the inference neural network $I$. The third term $\frac{\left\|\mathbf{z}\right\|^2}{2}$ in Equation \ref{eq:kl-objective} is the regularization term from the multivariate Gaussian prior which regularize the sampling close to the mean. 

As introduced in the previous section, it is difficult to evaluate the first term $\log q_{\phi}(\mathbf{z})$ in Equation \ref{eq:kl-objective} since it is unknown and intractable. It can be denoted as $-H(q_{\phi}) = \mathbb{E}_{\mathbf{z} \sim q_{\phi}} \left [\log q_{\phi}(\mathbf{z})\right ]$, where $-H(q_{\phi})$ is the negative entropy of $q_{\phi}$, it can be only evaluated through samples from the variational distribution. We use the Kozachenko-Leonenko estimator \citep{KozLeo87, Goria_etal_2005, chan2019parametric} to compute the estimation of the entropy of $q_{\phi}$ 

\begin{equation}
\hat{H}(q_{\phi}) = n_{\mathbf{z}} \cdot
\mathbb{E}_{\mathbf{z} \sim q_{\phi}} \log d_{k} + \textbf{c},
\vspace{+6pt}
\end{equation} 
where $d_{k}$ represents the distance between the latent vector $\mathbf{z}$ and its $k^{\text{th}}$ nearest neighbor from all the samples. We choose $k \approx \sqrt{M}$ according to \citep{Goria_etal_2005} and \citep{chan2019parametric}. The entropy estimator quantifies the spread of the sample elements. The larger the entropy means smaller $\mathbb{E}_{\mathbf{z} \sim q_{\phi}} \left [\log q_{\phi}(\mathbf{z})\right ]$. If the entropy term were excluded, minimizing Equation \ref{eq:kl-objective} would result in seeking a deterministic solution rather than sampling the entire posterior distribution \citep{chan2019parametric}.

Once the generator $G_{\theta}$ is trained, it remains unchanged. And to train the inference network $I_\phi$, we minimize the $\text{KL}\left[q_{\phi}(\mathbf{z}) \, \big\| \, p(\mathbf{z}|d_{\text{obs}}) \right]$ using gradient descent method. Commonly utilized optimization algorithms such as Adam or AdaGrad are employed for optimizing neural networks. The gradient of the loss with respect to the neural network weights $\phi$ is calculated through the adjoint state method and automatic differentiation using TensorFlow. The training process of $I_{\phi}$ is performing an inference, searching for the posterior distribution $P(\mathbf{z}|d_{\text{obs}})$, such that the inputs of the pretrained generator $G_{\theta}$ can generate subsurface images matches the seismic data. After the inference process, sampling conditional realizations is efficiently executed by directly sampling \( \mathbf{w} \sim P(\mathbf{w}) \) and conducting a forward pass through $ G_{\theta}(I_{\phi^*}(\mathbf{w})) $.

\subsection{Normalizing Flow}
\label{sec:flow}
In the previous section we introduced an implicit method to estimate the first term $\mathbb{E}_{\mathbf{z} \sim q_{\phi}} \left [\log q_{\phi}(\mathbf{z})\right ]$ in equation \ref{eq:kl-objective}. In machine learning statistics, a better estimation of probability density has been become the popular problem, but it is still hard. In deep learning, we use chain rule to back propagate the gradient to the trainable parameters, and the embedded probability distributions such as prior and posterior distributions are expected to be differentiated efficiently. That is why we normally start with Gaussian distribution in probabilistic models. However, real-world distributions can be more complex than just Gaussian distributions.

\begin{figure*}
    \centering
    \includegraphics[width=1\linewidth]{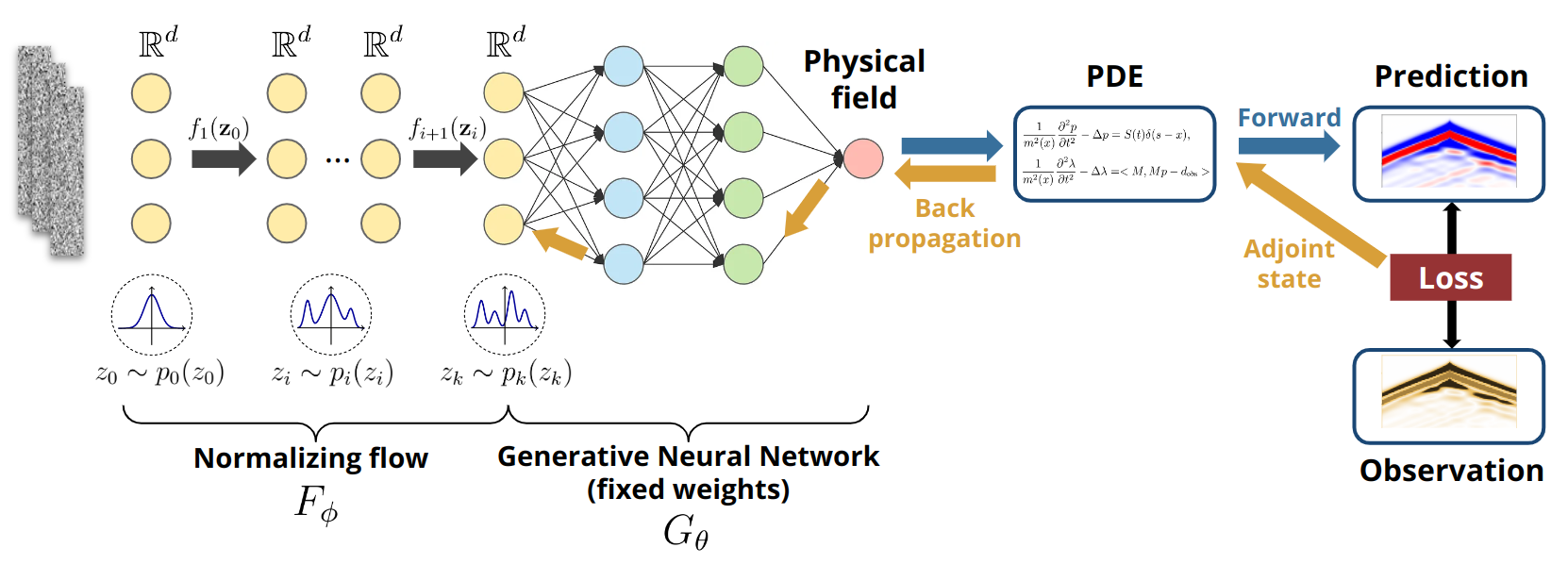}
    \caption{Structure of the explicit inference method using normalizing flow}
    \label{fig:flow-network}
\end{figure*}

Thus, we introduce a method starting from a Gaussian distribution to explicitly compute the pdf of the variational distribution $q_{\phi}$, we propose to use a normalizing flow $F_{\phi}$ to perform Bayesian inference and approximate the posterior distribution. After parameterizing of the subsurface image with GAN, we are able to perform sampling and conditioning in the latent space. As illustrated in Figure \ref{fig:flow-network}, our method performs Bayesian inference by attaching a normalizing flow-based inference network before the input of the pretrained generator, such that we can train the parameters of the normalizing flow and find an optimal approximation \( q_{\phi}^* (\mathbf{z}) \) to the posterior pdf \( p(\mathbf{z}|d_{\text{obs}}) \) within the family of known probability distributions \( Q = \{q_{\phi}(\mathbf{z}), \phi \in \mathbb{R}^{d} \} \).

A normalizing flow \citep{rezende2015variational} is considered to be capable to provide better and more powerful density estimation, it uses change of variable theorem for probability density estimation, starting from a random variable $\mathbf{z}_{i-1}$ and its known pdf $\mathbf{z}_{i-1} \sim p_{i-1}(\mathbf{z}_{i-1})$, normally Gaussian, to construct a new random variable $\mathbf{z}_i$ by using a one-to-one (bijective) mapping function $f$, hence $\mathbf{z}_i=f(\mathbf{z}_{i-1})$. And the function is invertible, so $\mathbf{z}_{i-1}=f^{-1}(\mathbf{z}_i)$. We can use the change of variable theorem:
% \begin{equation}
% p(\mathbf{z}_i) 
% = p(\mathbf{z}_{i-1}) \left\vert \det \dfrac{d \mathbf{z}_{i-1}}{d \mathbf{z}_i} \right\vert 
% = p_{i-1}(f_i^{-1}(\mathbf{z}_i)) \left\vert \det\dfrac{d f_i^{-1}}{d \mathbf{z}_i} \right\vert,
% \end{equation}

\begin{equation}
\begin{aligned}
\mathbf{z}_{i-1} &\sim p_{i-1}(\mathbf{z}_{i-1}) \\
\mathbf{z}_i &= f_i(\mathbf{z}_{i-1})\text{, thus }\mathbf{z}_{i-1} = f_i^{-1}(\mathbf{z}_i) \\
p_i(\mathbf{z}_i) 
&= p_{i-1}(\mathbf{z}_{i-1}) \left\vert \det \dfrac{d \mathbf{z}_{i-1}}{d \mathbf{z}_i} \right\vert = p_{i-1}(f_i^{-1}(\mathbf{z}_i)) \left\vert \det\dfrac{d f_i^{-1}}{d \mathbf{z}_i} \right\vert
\end{aligned}
\vspace{+6pt}
\end{equation}
where $\left\vert \det\dfrac{d f_i^{-1}}{d \mathbf{z}_i} \right\vert$ is the Jacobian determinant of the function $f$.
\vspace{+3pt}

Normalizing flow transforms a simple distribution into a complex one by applying the change of variable theorem using a sequence of invertible transformation functions. We infer new distributions by transforming from the previous one, and the final distribution is transformed many times and is expected to be able to approximate the complex target posterior distribution.

% , as is shown in Figure \ref{fig:flow}.

% \begin{figure}
%     \centering
%     \includegraphics[width=0.5\linewidth]{images/flow/normalizing-flow.png}
%     \caption{Normalizing flow with transformation function $f_i$}
%     \label{fig:flow}
% \end{figure}

And it is notable that $z$ and $z_{i-1}$ need to be continuous and have the same dimension, this is made possible through dimensional reduction using GAN. Because at each transformation process of a normalizing flow, we need to compute the inverse of the function $f_i^{-1}(\mathbf{z}_i)$ and the Jacobian determinant $\left\vert \det\dfrac{d f_i^{-1}}{d \mathbf{z}_i} \right\vert$, the transformation function is expected to match two conditions:
\begin{enumerate}
    \item Easily invertible
    \item Jacobian determinant is easy to compute
\end{enumerate}

The Real Non-Volume Preserving (RealNVP) model is such method composes two kinds of invertible transformations: additive coupling layers and rescaling layers. It uses a translation $t$ and scaling factors $s$ to the transformation:

\textbf{Forward Mapping:}
\begin{itemize}
    \item $x_1=z_1$, which represents an identity mapping.
    \item $x_2=\exp (s(z_1))\odot z_2+t(z_1)$
\end{itemize}

\textbf{Inverse Mapping:}
\begin{itemize}
    \item $z_1=x_1$, which represents an identity
    \item $z_2=(x_2-t(x1)) \odot \exp(-s(x1))$
\vspace{+6pt}
\end{itemize}
where $\odot$ denotes elementwise product. This results in a non-volume preserving transformation. It is easily invertible according to the properties of the inverse mapping. Since both inverting the transform function $f$ and computing the determinant does not require does not require computing the inverse of $s$ and $t$, these two functions can complex and be parameterized by weights of invertible neural networks. And the normalizing flow can be optimized by applying the gradient of objective function with respect to the trainable weights of $s$ and $t$. 

% The mapping process is shown in figure \ref{fig:realnvp}

% \begin{figure}
%     \centering
%     \includegraphics[width=1\linewidth]{images/realnvp.png}
%     \caption{RealNVP model}
%     \label{fig:realnvp}
% \end{figure}

And the Jacobian matrix is given by a lower-triangular matrix:
\begin{equation}
\mathbf{J} = 
\begin{bmatrix}
  \mathbb{I}_d & \mathbf{0}_{d\times(D-d)} \\[5pt]
  \frac{\partial \mathbf{y}_{d+1:D}}{\partial \mathbf{x}_{1:d}} & \text{diag}(\exp(s(\mathbf{x}_{1:d})))
\end{bmatrix}
\end{equation}

Hence the determinant of the Jacobian matrix is easy to get by computing the product of the diagonal items.

\begin{equation}
\det(\mathbf{J}) 
= \prod_{j=1}^{D-d}\exp(s(\mathbf{x}_{1:d}))_j
= \exp(\sum_{j=1}^{D-d} s(\mathbf{x}_{1:d})_j)
\end{equation}

Thus, the first term $\mathbb{E}_{\mathbf{z} \sim q_{\phi}} \left [\log q_{\phi}(\mathbf{z})\right ]$ in Equation \ref{eq:kl-objective} can be explicitly measured by a set of transformation functions $f_i$ of the normalizing flow $F_{\phi}$ using the change of variable theorem.

\begin{equation}
z = f_n \circ f_{n-1} \circ \ldots \circ f_2 \circ f_1 (x)
\end{equation}

\subsection{Stein Variational Gradient Descent (SVGD)}
SVGD is a variational technique that employs a set of samples, known as particles, to approximate the pdf of \( q_{\phi}(\mathbf{z}) \). These particles are iteratively refined to minimize the Kullback-Leibler (KL) divergence, aligning the final particle distribution with the posterior distribution \citep{liu2016stein}. By representing the variational distribution with particles, SVGD provides flexibility in exploring posterior distributions with complex shapes. We represent the set of particles as \(\{\mathbf{z}_i\}\), where each \(\mathbf{z}_i\) is a particle in a \(d\)-dimensional latent space. SVGD applies a smooth transformation to update these particles.

\begin{equation}
T(\mathbf{z}) = \mathbf{z} + \epsilon \phi^*(\mathbf{z}),
\end{equation}
where \( \phi = [\phi_1, \ldots, \phi_d] \) is a smooth vector function indicating the direction of perturbation, and \( \epsilon \) is the step size. Assuming \( T \) is invertible, the transformed pdf \( q_T (\mathbf{z}) \) can be defined. The gradient of the KL divergence between \( q_T \) and the posterior pdf \( p \) with respect to \( \epsilon \) is given by \citep{liu2016stein}:

\begin{equation}
\nabla_\epsilon \text{KL}[q_T \| p]\bigg|_{\epsilon=0} = -\mathbb{E}_{q} [\text{trace} (\mathcal{A}_p \phi(\mathbf{z}))],
\end{equation}
where \( \mathcal{A}_p \) is the Stein operator, defined as \( \mathcal{A}_p \phi(\mathbf{z}) = \nabla_\mathbf{z} \log p(\mathbf{z}) \phi(\mathbf{z})^T + \nabla_\mathbf{z} \phi(\mathbf{z}) \). By maximizing this expectation, we find the steepest descent direction of the KL divergence. Minimizing the KL divergence can thus be achieved by iteratively stepping in this direction.

The optimal perturbation direction \( \phi^* \) can be determined using kernel functions. For \( x, y \in X \), let \( \psi \) be a mapping from \( X \) to a Hilbert space where an inner product \( \langle , \rangle \) is defined. A kernel function \( k(x, y) \) satisfies \( k(x, y) = \langle \psi(x), \psi(y) \rangle \). With a kernel \( k(\mathbf{z}', \mathbf{z}) \), the optimal \( \phi^* \) is expressed as \citep{liu2016stein}:

\begin{equation}
\phi^* \propto \mathbb{E}_{\mathbf{z}' \sim q} [\mathcal{A}_p k(\mathbf{z}', \mathbf{z})].
\end{equation}

In practice, since particles \(\{\mathbf{z}_i\}\) represent \( q \), the expectation can be approximated by the particle mean. The KL divergence is minimized by iteratively applying the transform \( T(\mathbf{z}) = \mathbf{z} + \epsilon \phi^*(\mathbf{z}) \) to a set of initial particles \(\{\mathbf{z}_i^0\}\):

\begin{equation}
\phi_l^*(\mathbf{z}) = \frac{1}{n} \sum_{j=1}^n \left[ k(\mathbf{z}_l^j, \mathbf{z}) \nabla_{\mathbf{z}_l^j} \log p(\mathbf{z}_l^j|d_{\text{obs}}) + \nabla_{\mathbf{z}_l^j} k(\mathbf{z}_l^j, \mathbf{z}) \right],
\label{eq:svgd_grad}
\end{equation}

\begin{equation}
\mathbf{z}_{l+1}^i = \mathbf{z}_l^i + \epsilon_l \phi_l^*(\mathbf{z}_l^i),
\label{eq:svgd_update}
\end{equation}
where \( l \) is the iteration index, \( n \) is the number of particles, and \( \epsilon_l \) is the step size. If the step size \( \{\epsilon_l\} \) is sufficiently small, the transform \( T \) remains invertible, and the process asymptotically converges to the posterior pdf as the number of particles increases. A common kernel function used is the radial basis function (RBF):

\begin{equation}
k(\mathbf{z}', \mathbf{z}) = \exp\left[ -\frac{||\mathbf{z} - \mathbf{z}'||^2}{2h^2} \right],
\end{equation}
where \( h \) is a scale factor that controls the interaction intensity between particles based on their distance. Following \citep{liu2016stein}, \( h \) is chosen as \( h = d_{\tilde{}} / \sqrt{2 \log n} \), where \( d_{\tilde{}} \) is the median of pairwise distances between all particles. This ensures balanced interaction among particles, avoiding excessive clustering or dispersion:
\begin{equation}
\sum_{j \neq i} k(\mathbf{z}_i, \mathbf{z}_j) \approx n \exp\left(-\frac{1}{2h^2} \tilde{d_{\tilde{}}}^2\right) = 1.
\end{equation}

For the RBF kernel, the second term of \( \phi^* \) in the gradient equation becomes:
\begin{equation}
\sum_j \frac{(\mathbf{z} - \mathbf{z}_j)}{\sigma^2} k(\mathbf{z}_j, \mathbf{z}),
\end{equation}
which acts as a repulsive force preventing particles from collapsing to a single mode. The first term consists of kernel-weighted gradients driving particles toward high-probability areas.

SVGD is an efficient particle-based variational Bayesian method for sampling from complex posterior distributions. By iteratively refining a set of particles to approximate the target posterior, SVGD strikes a balance between accuracy and computational efficiency by representing the distribution using a set of particles, leveraging gradient information for Bayesian inference.

\section{Results}
Following the training of the GAN, we employed the mentioned inference methods in a conditional generation setup. We compute the gradient of the KL divergence objective function, and numerical computations of solving adjoint state equation are executed utilizing GPU acceleration. PDE is solved numerically using second-order finite difference scheme. The true velocity model is from the original Overthrust dataset, the computational grid comprises 40-depth ($z$) by 120-width ($x$) grids.

\begin{figure*}
\centering
\includegraphics[width=0.75\textwidth]{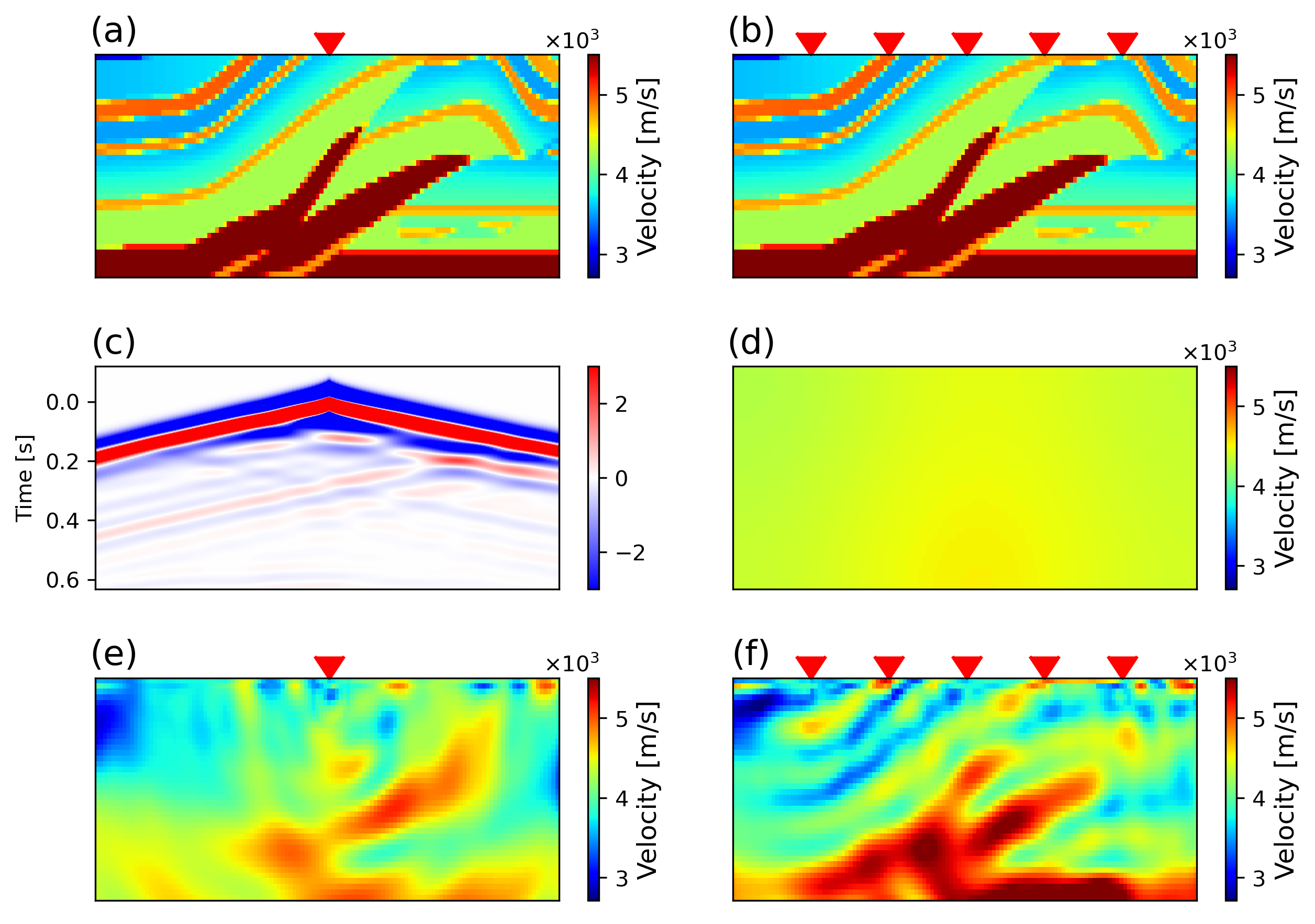}
\caption{Visualization of the velocity field depicting the subsurface structure, with receivers distributed across every grid point on the surface of the model. \textbf{a} The configuration with one seismic source, while \textbf{b} displays the setup with five seismic sources positioned on the model surface. \textbf{c} The simulated seismic shot gather. \textbf{d} The initial guess for standard FWI. The inversion results corresponding to \textbf{e} one source and \textbf{f} five sources}
\label{fig:forward}
\end{figure*}

In our numerical experiments, we solve the acoustic wave equation using a frequency cap of 25Hz, processing 600 iterations in the time domain with a discrete interval of 1ms. The spatial grid points, set at $9.5 \text{m}$ intervals for both 
$dz$ and $dx$, and with grid points spaced at intervals of 8.71 meters in both the z and x directions. Receivers are distributed across every grid point on the surface, with different source configurations on the surface of the model. Figure \ref{fig:forward} showcase the numerical simulation obtained using one source and five sources configuration, Figure \ref{fig:forward}c showcase the seismic shot gather of forward modeling using one source configuration. Figure \ref{fig:forward}d shows the initial guess of velocity model used as the starting point for standard FWI. Figures \ref{fig:forward}e and f show the standard FWI results obtained with configurations using one source and five sources, respectively. With the one-source configuration, the inversion result exhibits a lack of information at the lower corners of the model. Conversely, the five-source configuration provides a better reconstruction of the layers in the upper part of the model and the split Overthrust structure in the central region. However, there is still a deficiency of information at the lower corner locations compared to the lower central part. It is noteworthy that we utilize the L-BFGS-B algorithm \citep{liu1989limited, zhu1997algorithm} for the optimization of standard FWI. This algorithm incorporates boundary conditions during the optimization process. However, after parameterization using GAN, the minimum and maximum velocities are already constrained within the bounds of the training dataset from the Overthrust model, this implies that all generated images from GAN fall within the parameter limits. 

We assume that a Gaussian distribution can represent the noise of the seismic data in the likelihood function

\begin{equation}
P(d_{\text{obs}}|\mathbf{z}) \propto \exp\left(\frac{-\left\| d_{\text{pred}} - d_{\text{obs}} \right\|^2}{2\sigma^2}\right) = \exp\left(-\frac{J(\mathbf{z})}{\sigma^2}\right),
\vspace{+6pt}
\end{equation}
where $\sigma$ represents the standard deviation of the data, and it the parameter $\sigma$ in the likelihood function influences the weighting or strength of belief assigned to the prior term. In the context of GAN-regularized FWI, a larger value of $\sigma$ results in the latent vector being sampled closer to the Gaussian mean in latent space, thereby enhancing the realism of the generated images produced by the GAN. In this study, we initially set $\sigma = 10^2$ as the regularization for the exploration in the latent space during optimization. Subsequently, we can progressively reduce the value of sigma to mitigate noise and refine the model convergence. Both inference methods eventually allows us to conditionally generate seismic images consistent to the synthetic reference seismic data. 

For inference network, we choose initial neural network input distribution $p_{\mathbf{w}}$ as a multivariate Gaussian distribution, and the output of the inference network has the same dimensionality $d_{\mathbf{w}} = d_{\textbf{z}} = 100$, which is the input of the generator. We choose the batch size of 32, and we use automatic differentiation to compute the gradient of the KL-divergence with respect to the trainable parameters of the inference neural network $I_\phi$, and we apply the gradient to update $I_\phi$ using the Adam optimization algorithm \citep{kingma2014adam} with a learning rate of $10^{-5}$. We update 1000 iterations until the average data residual of the posterior samples ceases to decrease. After the generator finishes training, it 
remains fixed and only provides gradient backpropagation during inference, which does not require any extra computation. Post-inference, generating samples from the posterior distribution is highly efficient. It involves simply sampling latent vectors from $\textbf{w} \sim \mathcal{N}(0, I^{100})$ and passing them through two neural networks $G_\theta(I_{\phi^*}(\textbf{w}))$.

For the explicit inference method using normalizing flow involves initializing a series of bijectors and applying them to transform the distribution. The flow consists of 20 layers, where each layer comprises an autoregressive network with two hidden layers of 512 units each and ReLU activation. We choose the batch size of 32, the normalizing flow is optimized using the Adam optimizer with a learning rate of $5 \times 10^{-4}$. The trainable parameters of normalizing flow $F_{\phi}$ are then updated for 1000 iterations until the average data residual of the posterior samples ceases to decrease.

For SVGD, we use 512 particles sampled from a 100-dimensional Gaussian prior distribution. We use automatic differentiation computes the particle updates at each step in equation \ref{eq:svgd_grad}, and we update all particles with the smooth transform using equation \ref{eq:svgd_update}. We use Adam optimizer updates the particles with a learning rate of $5 \times 10^{-1}$ for 100 iterations, until the average data residual of the posterior samples stops decreasing. The final 512 particles in the latent space generate posterior samples conditioned on the observations.

We perform McMC sampling according to the MALA algorithm. We initialize 20 MALA chains with initial step size of $\epsilon_{t=0}= 10^{-3}$. To ensure stable samples in the posterior distribution, we gradually decrease the step size until $\epsilon_{t=100}= 10^{-5}$ \citep{mosser2020stochastic}. After 100 iterations, we maintain the step size unchanged for the remaining sampling process. After the completion of the sampling, we choose a burn-in period of 50 iterations for the MALA method.

Note that all inference methods used in this study require solving the PDE twice using the adjoint state method for each computation of the gradient of the negative log likelihood function. This step is the primary challenge of the inference process. Compared to the numerical computation above, differentiating the neural networks is more computationally efficient. Additionally, it is worth mentioning that training the GAN or other types of generative models is generally more demanding compared to the subsequent inference process.

\subsection{Model comparison}

\begin{figure*}
\centering
\includegraphics[width=0.75\textwidth]{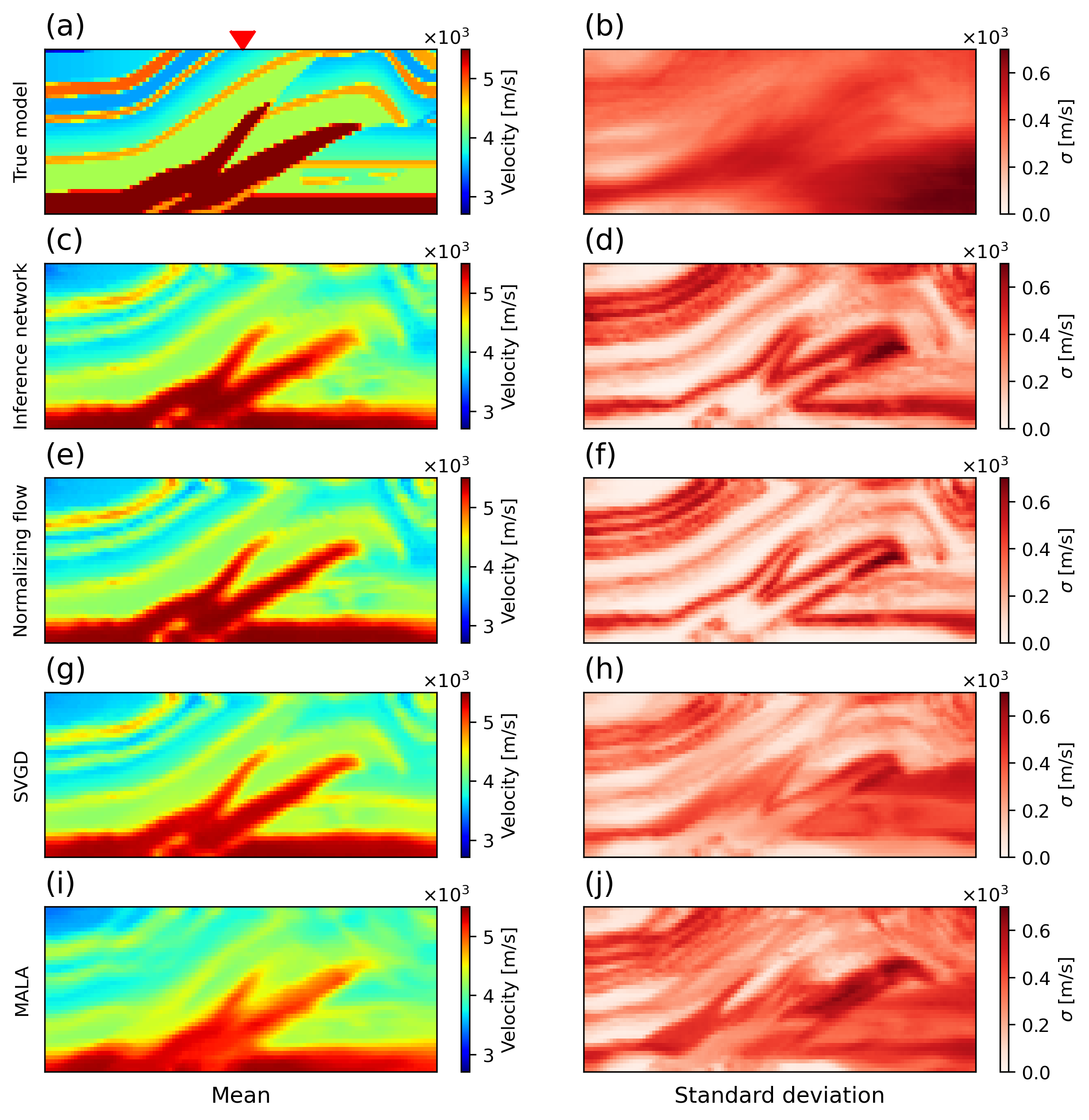}
\caption{Model comparison between inference Network, normalizing flow, SVGD and MALA. \textbf{a} True velocity model with the source location marked by a red triangle at the surface center. \textbf{b} Prior standard deviation representing uncertainty inferred from GAN-generated distributions. \textbf{c, d} Mean and standard deviation obtained via the inference network. \textbf{e, f} Mean and standard deviation derived from normalizing flow. \textbf{g, h} Mean and standard deviation computed using the SVGD method. \textbf{i, j} Mean and standard deviation computed using the MALA method.}
\label{fig:model_comparison}
\end{figure*}

The model comparison between inference network, normalizing flow, SVGD and MALA are showcased in Figure \ref{fig:model_comparison}. These methods are assessed based on their ability to reconstruct the true velocity model and quantify uncertainties. 

Figure \ref{fig:model_comparison}a illustrates the true velocity model, with a red triangle marking the source location positioned at the center of the surface. Figure \ref{fig:model_comparison}b displays the prior standard deviation, indicating the uncertainty inferred from the GAN-generated distributions. Figure \ref{fig:model_comparison}c, d present the mean and standard deviation obtained via the inference network. Figure \ref{fig:model_comparison}e, f depict the mean and standard deviation derived from normalizing flow. Figure \ref{fig:model_comparison}g, h exhibit the mean and standard deviation computed from the SVGD particles. Figure \ref{fig:model_comparison}i, j exhibit the mean and standard deviation computed from the MALA samples after the burn-in period. The examination of the results reveals that all four methods reproduce the true velocity model, particularly capturing the split structure characteristic of the Overthrust using a single source. In regions featuring sloping structures, there is a pronounced presence of high uncertainties due to the positioning of the largest contrast, visible at the right end of Overthrust structure.

In comparison, the inference network and normalizing flow methods exhibit smaller uncertainties than SVGD and MALA. This observation indicates the use of neural network to transform a simple distribution such as a multivariate Gaussian, can provide very limited capability to transform it to an arbitrarily complex target distribution, neither using entropy estimator nor change of variable theorem, which inherently provides less variability compared to the SVGD and MALA method which use particles or desecrate samples to explore the posterior. The complexity of seismic inversion in high-dimensional spaces poses computational challenges to reconstruct the continuous full posterior distribution. The neural network based methods using inference network or normalizing flow tend to fall into one mode that generate samples that are close to each other, thereby diminishing uncertainties of multimode problems in the high-dimensional space. Additionally, despite the expected lower information in the deeper part of the model, all methods consistently show low uncertainties in this region where lack of information from the seismic observations. This contradiction can be attributed to the high constraint imposed by the prior distribution well-learned by the GAN in this specific part, as all training images exhibit high velocities with relatively low variance in this region, limited by the poverty of the training images.

Although the lower corners of the model contain limited information from recorded seismic waveforms, the high correlation of the grids close to the edge with their neighboring points (Figure \ref{fig:correlation_prior}f) which exhibits in the prior distribution facilitates the automatic extension of information to regions with minimal to no information in standard FWI. This experiment was conducted with only one source yet achieved successful reconstruction of the true velocity model, attributed to the GAN provision of a smaller search range. 

The results of posterior mean using all methods presented in Figure \ref{fig:model_comparison} shows the capability of proposed stochastic methods in capturing the complexities of subsurface structures while accounting for uncertainties inherent in seismic inversion.

\begin{figure*}
\centering
\includegraphics[width=0.7\textwidth]{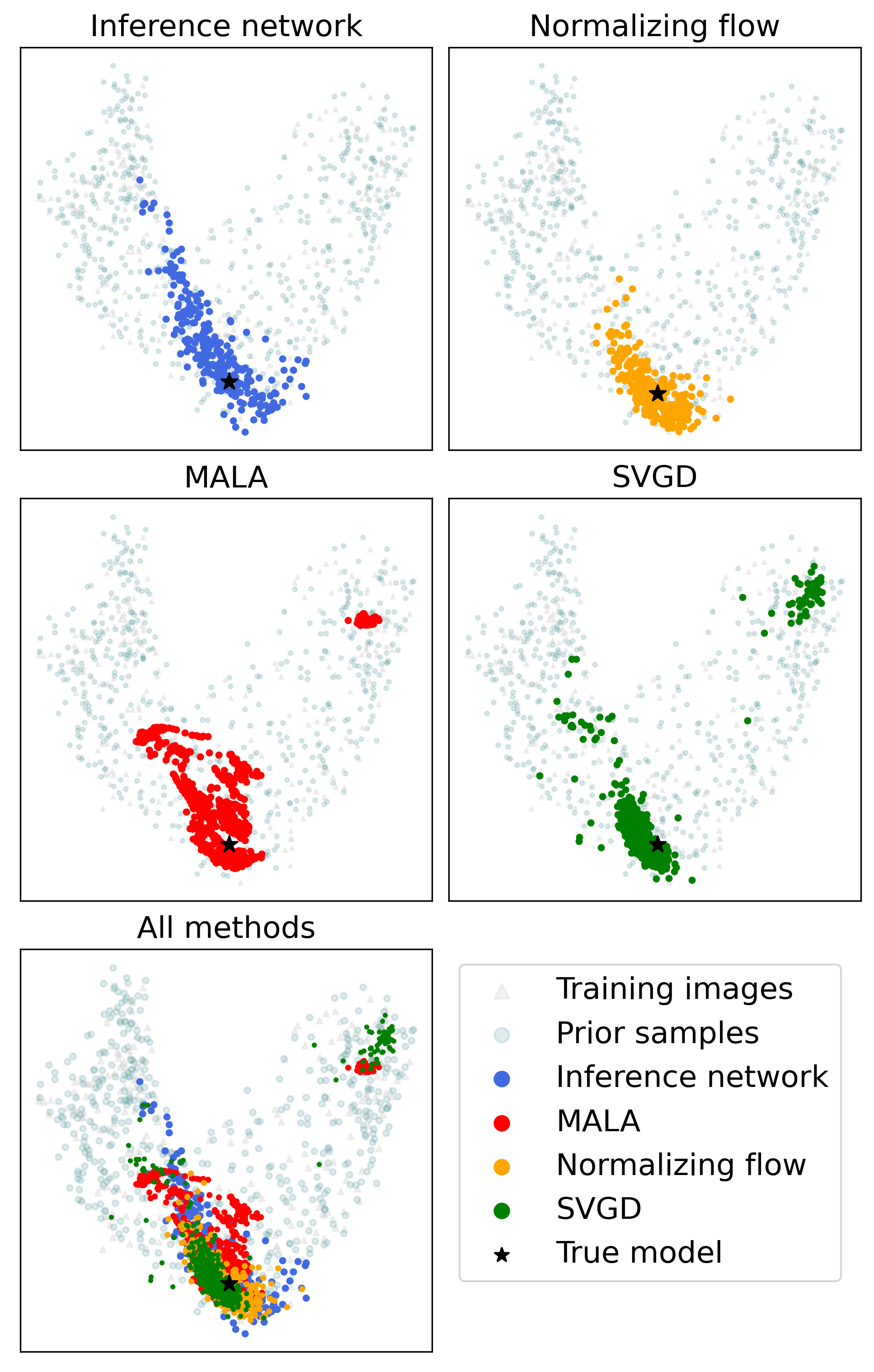}
\caption{PCA visualization of data points in 2D representing the posterior samples obtained from inference network, normalizing flow, SVGD and MALA methods.}
\label{fig:pca_comparison}
\end{figure*}

Figure \ref{fig:pca_comparison} showcases the visualization in 2D by projecting all the images to the two principal components of the GAN-generated image space using principal component analysis (PCA), showcasing the distribution of images in a reduced feature space. Furthermore, the samples obtained using normalizing flow, inference network, SVGD and MALA are displayed in the same two-dimensional space. The results reveal variations in the distribution sizes among these methods. The largest range is observed for the training images and prior distribution learned by the GAN as expected. For the posterior distribution obtained by different methods, the inference network provides a larger posterior distribution size compared to the samples obtained by normalizing flow, while both methods have limited flexibilities and only address a single mode as the final solution.

While the discrete posterior sampling methods SVGD and MALA provide more variability, this trend aligns with earlier findings. SVGD and MALA tend to move particles or chains into different modes, resulting in greater variability in posterior samples. In contrast, both neural network-based variational Bayesian inference methods show less variability, with the normalizing flow exhibiting a more regularly shaped posterior distribution than the inference network method. Notably, in the reduced-dimensional space, the true model is located within the distribution of all four methods, indicating that each method includes the correct true model in its posterior distribution. Additionally, several SVGD particles are distant from the main clusters of other particles, capturing modes that are not identified by other methods.

\begin{figure*}
\centering
\includegraphics[width=0.75\textwidth]{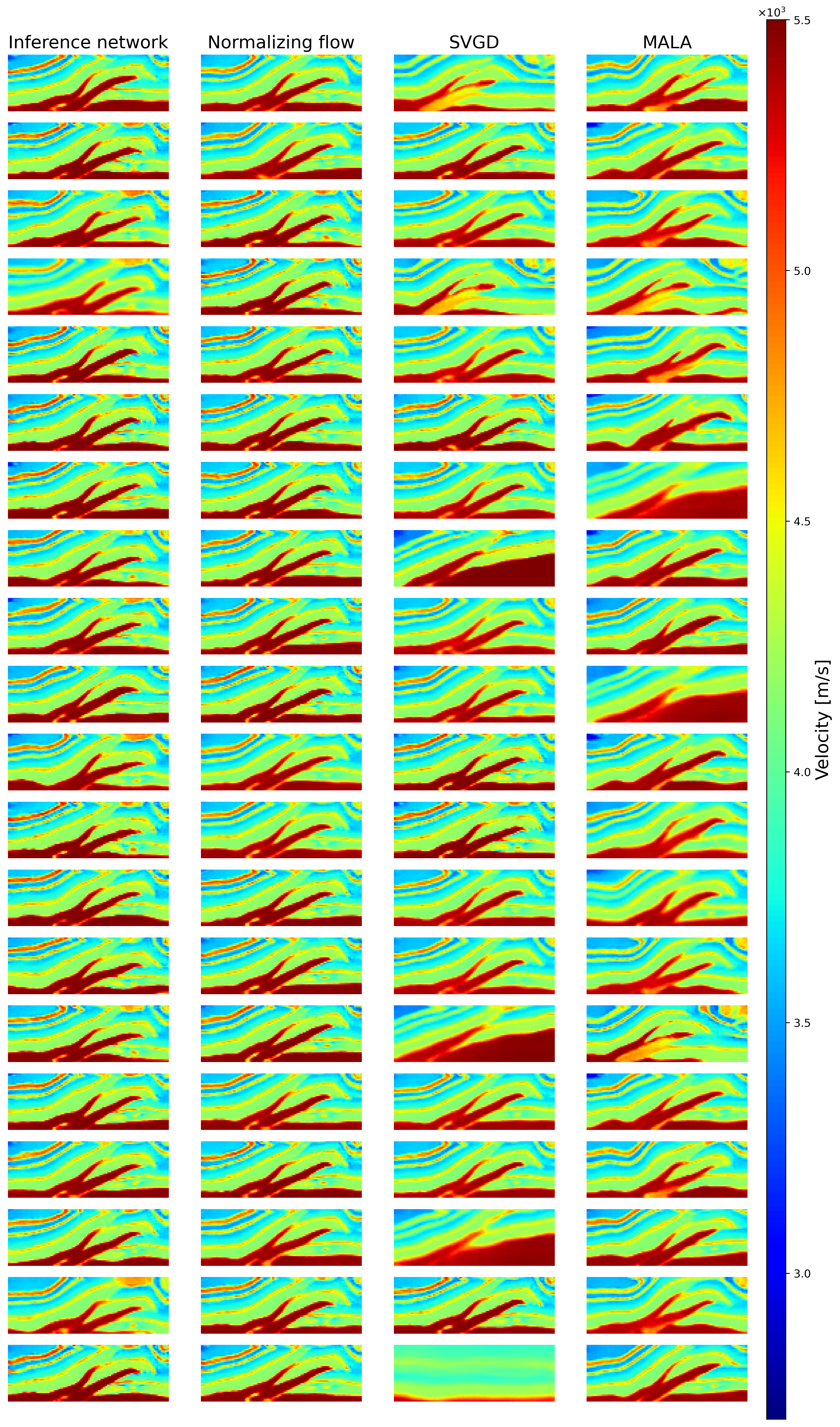}
\caption{Comparison of realizations generated from the posterior distribution using inference network, normalizing flow, SVGD and MALA}
\label{fig:comparison_samples}
\end{figure*}

\begin{figure*}
\centering
\includegraphics[width=0.8\textwidth]{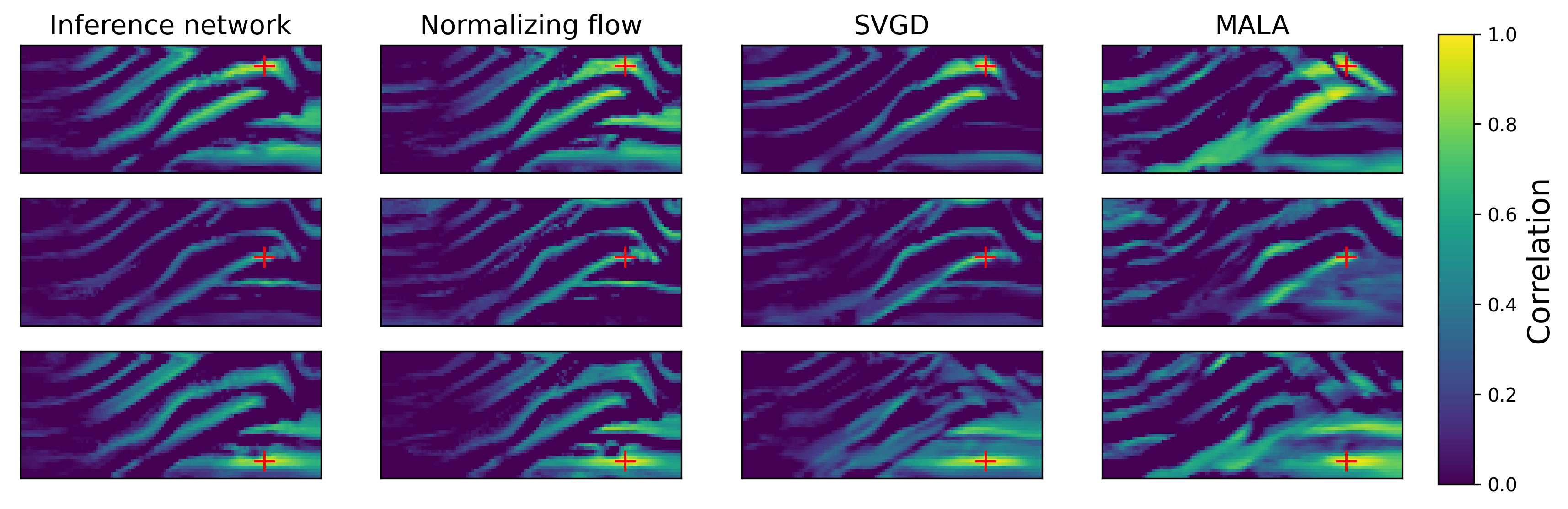}
\caption{Comparison of correlation at three different locations in the velocity field using the posterior sampled from inference network, normalizing flow, SVGD and MALA. The cross indicates the grid point used to compute the correlation.}
\label{fig:comparison_correlation}
\end{figure*}

Figure \ref{fig:comparison_samples} showcases a comparison of realizations generated from the posterior distribution using four methods: inference network, normalizing flow, SVGD and MALA. Although the samples generated by the inference network and normalizing flow are similar, they exhibit a tendency towards images close to the true model. This observation aligns with the PCA visualization shown in Figure \ref{fig:pca_comparison}, where all posterior samples derived using the inference network and normalizing flow are within a single mode. In contrast, the velocity fields generated by SVGD and MALA display more variability, indicating their ability to provide different images that match the seismic observations and better address uncertainties in high-dimensional space. All these methods demonstrate similar texture smoothness by using the same noise level $\sigma = 10^2$ in the likelihood function, thereby enhancing the overall quality of the generated samples. If we reduce $\sigma$, it may increase the flexibility of the sampling, leading to more variations but also potentially introducing structural artifacts, such as unrealistic textures.

Figure \ref{fig:comparison_correlation} illustrates the comparison of correlation at four different depths in the velocity field using samples from the posterior distributions obtained via the inference network, normalizing flow, SVGD and MALA. At the top location, MALA exhibits higher correlation along the Overthrust structure compared to the other methods. Overall, all methods demonstrate higher correlation at the grids surrounding the Overthrust structure, indicating a well reconstruction of the true model. Additionally, all methods show high horizontal correlation at deeper locations, extending the knowledge from the central part to the model edges, which is attributable to the nature of the prior distribution.

\begin{figure*}
\centering
\includegraphics[width=0.7\textwidth]{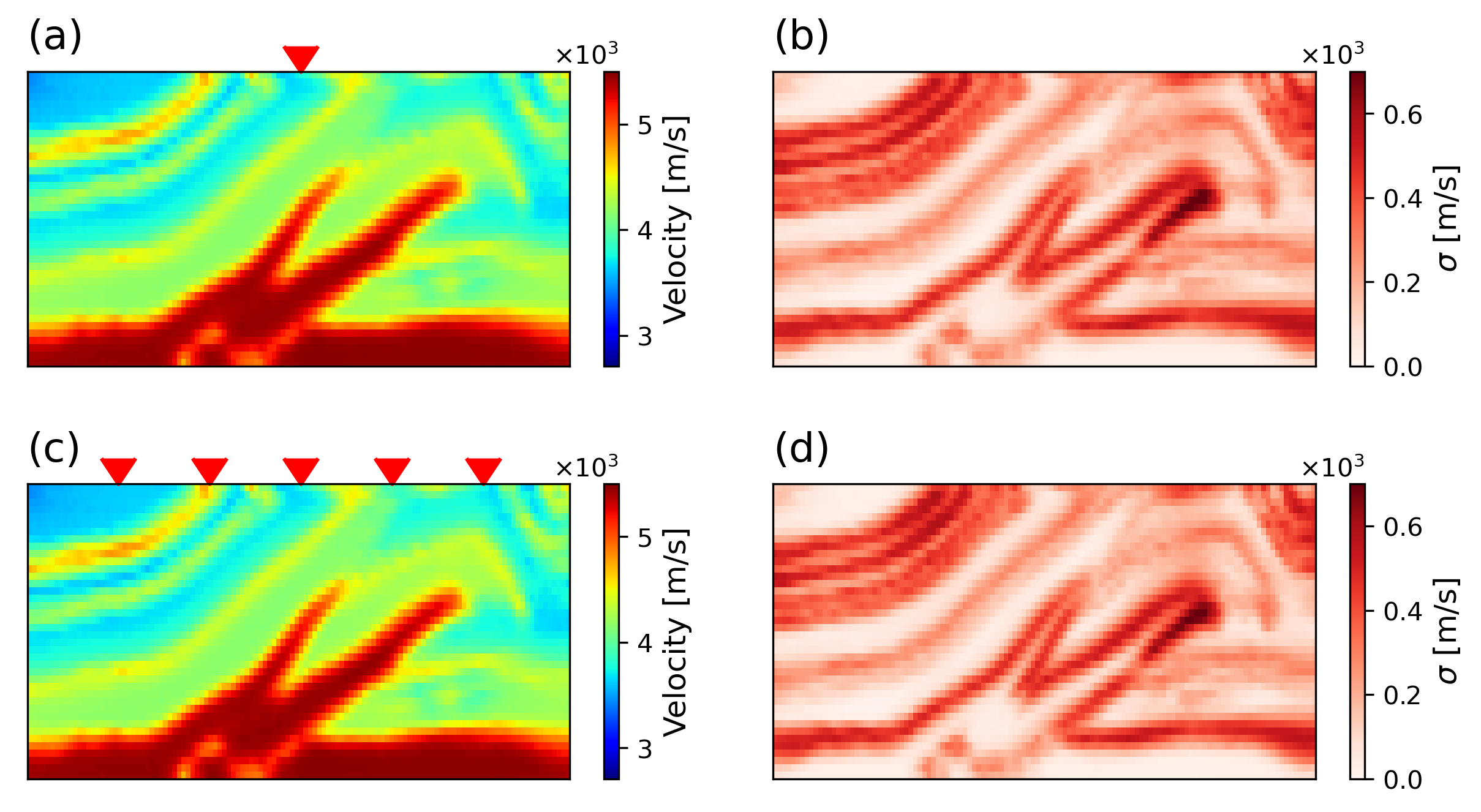}
\caption{Comparison of the results obtained using normalizing flow with different numbers of sources. \textbf{a}, \textbf{b} The posterior mean and uncertainties, respectively, for the configuration with one seismic source. \textbf{c} , \textbf{d} The posterior mean and uncertainties, respectively, for the configuration with five seismic sources.}
\label{fig:source_comparison}
\end{figure*}

% \begin{figure*}
% \centering
% \includegraphics[width=0.7\textwidth]{images/comparison/source_difference.png}
% \caption{Comparison of the results obtained using normalizing flow with different numbers of seismic sources. \textbf{a} The difference between the posterior mean velocity fields ($V_{\text{mean}}$) for one and five sources ($n_{\text{source}}=1$ and $n_{\text{source}}=5$) configurations. \textbf{b} The difference between the posterior standard deviations ($\sigma_{\text{post}}$) for one and five sources ($n_{\text{source}}=1$ and $n_{\text{source}}=5$) configurations.}
% \label{fig:source_difference}
% \end{figure*}

\begin{figure*}
\centering
\includegraphics[width=0.7\textwidth]{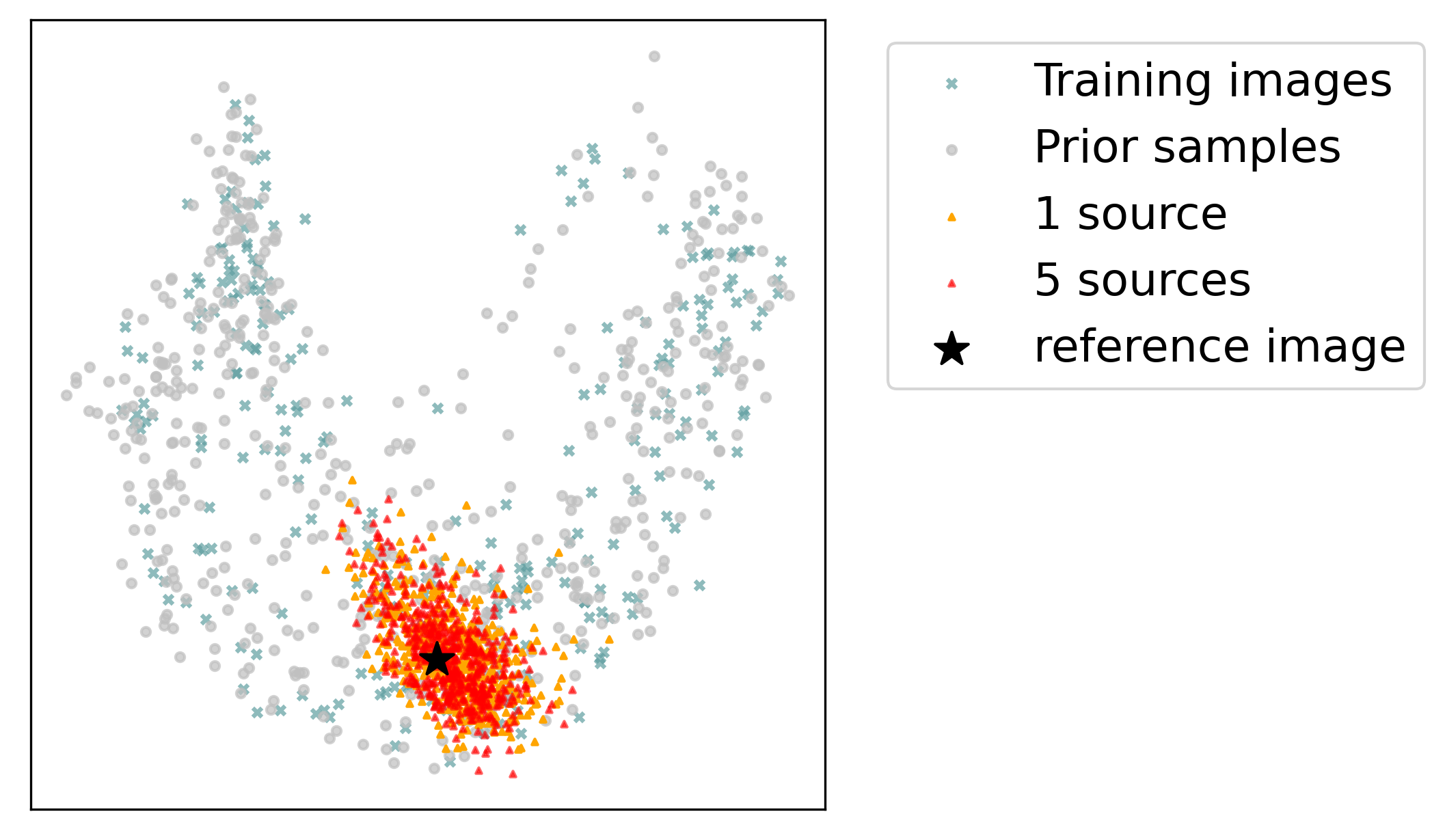}
\caption{PCA visualization of data points obtained using normalizing flow with 1 source and 5 sources.}
\label{fig:source_pca}
\end{figure*}

Figure \ref{fig:source_comparison} compares the results obtained using normalizing flow with different numbers of seismic sources. In Figure \ref{fig:source_comparison}a, we visualize the difference between the posterior mean velocity fields ($V_{\text{mean}}$) for one and five sources configurations, Figure \ref{fig:source_comparison}b shows the difference between the posterior standard deviations ($\sigma_{\text{post}}$) for the same configurations. Figure \ref{fig:source_comparison}a indicates that the mean velocity difference mainly exists in the edges of the Overthrust structure, while Figure \ref{fig:source_comparison}b suggests that the one source configuration exhibits slightly higher uncertainties at the lower corner region. Figure \ref{fig:source_pca} showcases the PCA visualization of 512 data points obtained using normalizing flow with 1 source and 5 sources, respectively. The PCA result reveals that the stochastic inversion results obtained with 1 source and 5 sources exhibit a similar range of posterior distribution. However, the 1 source configuration shows several points drifting away from the majority of the points. These observations demonstrate the capability to utilize a sparse source configuration while incorporating a GAN as the prior distribution.

\subsection{Computational cost}
We conducted our computations using a NVIDIA 1080 GPU with 8GB of memory (2560 CUDA Cores) and utilized the CuPy library for numerical calculations of the wave equation. The gradient of the data residual was obtained using the adjoint wave equation. To expedite computations in Python and CuPy, we adopted a matrix-based approach, and the acoustic wave equation (Equation \ref{eq:acoustic}) can be written as
\begin{equation}
p^{(t+1)} = 2p^{(t)} + \Delta p - p^{(t-1)} + S(t)\delta(x - s).
\end{equation}
where Laplacian $\Delta p$ is the function of $p^{(t)}$. In order to present this numerically tractable, the wavefield update equation can be cast in a matrix-based form
\begin{equation}
p^{(t+1)} = A \cdot p^{(t)} - p^{(t-1)} + src,
\end{equation}
where $A$ represents a transition matrix that operates on the current wavefield $p^{(t)}$, to transform from the wavefield $p^{(t-1)}$ at the previous time step, and $\text{src}$ is the contribution from the source term.

In stochastic seismic imaging, the most computationally intensive task is obtaining the gradient of the data residual in the likelihood function. The time required is directly proportional to the number of sources, without parallelization on a single GPU. In our study, the stochastic gradient descent using inference network with a batch size of 32 requires 1000 iterations (308 minutes) to reach a stable point where the KL divergence did not decrease further. In contrast, normalizing flow  has faster convergence during optimization with a training batch size of 32, requiring 200 iterations (182 minutes) to achieve satisfactory results. However, while normalizing flow converged faster, each iteration took longer due to the more complex structure involved in passing through the transformation function. For SVGD, the time required depends on the number of particles used. For example, in our experiment, SVGD with 512 particles costs 287 minutes. It is noteworthy that the computation for the SVGD kernel function $k$ in equation \ref{eq:svgd_grad} can be time-consuming depending on the number of particles. For MALA, due to its inherent nature of MCMC sampling, the time required is proportional to the number of samples needed. For instance, running MALA with 20 chains for 150 iterations to obtain 3000 samples (including the burn-in period) costs 231 minutes in our study. 

Considering the ability of SVGD and MALA to provide ideal variability in the final posterior distribution, along with the lower efficiency of the inference network and normalizing flow methods, SVGD and MALA emerge as the preferred choices for stochastic FWI with GAN as the prior. While the discrete posterior sampling methods, SVGD and MALA offer greater variability in the results, the neural network-based variational methods using the inference network and normalizing flow enable the generation of a continuous posterior distribution, from which an unlimited number of posterior samples can be simulated after the inference process.

However, we note that discussions about efficiency and computational costs may not be entirely accurate due to hardware variations, the efficiency of programming languages, the batch size, the number of particles and the number of inference iterations. Moreover, since solving the adjoint wave equation to obtain the gradient is the most time-consuming part of the computation, the comparison is most relevant to computational inverse problems where obtaining the gradient is challenging.

\section*{Discussion}

Our study provides an examination of four inference methods for stochastic seismic imaging: inference network, normalizing flow, SVGD and MALA using a GAN generator as the prior knowledge of the subsurface structures. Through extensive experimentation, we reveal the strengths and weaknesses of each approach, giving insight into their effectiveness in capturing subsurface structures and quantifying uncertainties.

Our findings underscore the power of deep generative models, particularly GAN, in learning complex distributions of subsurface velocity fields. By using GAN as Bayesian priors, we show its ability to generate synthetic seismic images that closely resemble real observations, offering valuable insights into the subsurface structure. Integrating variational Bayesian inference methods which also include deep neural networks, we enable uncertainty quantification and providing a novel framework for stochastic seismic inversion and computational inverse problems.

Comparing our results obtained using all methods. By implicitly or explicitly computing the density of the variational distribution, the inference network and normalizing flow offer alternative approaches in dimensional reduced latent spaces. Both methods show the capability to obtain the continuous posterior distribution and efficient post-training sampling and unlimited number of image generation from the posterior distribution. However, both methods provide small variability and fail to address all modes. It is notable that the neural network has limited capability to transfer a simple Gaussian distribution into a rather complex distribution, it struggles to capture the full complexity of subsurface structures and quantify uncertainties effectively. The efficiency in high dimensional space makes SVGD and MALA are suitable for large-scale seismic imaging and other computational inverse problem applications. These methods utilize a few particles or samples to represent the distribution, potentially improving efficiency for high-dimensional problems. While the set of particles or samples may not fully represent the posterior distribution, they can offer reasonable estimates of mean and standard deviation. 

Additionally, the computational demands of training GANs pose a challenge. Despite using a small dataset from Overthrust and employing data augmentation, expanding the scope of features necessitates a larger dataset, consuming more computational resources and time. Future research can focus on more efficient architectures to enhance the training process of generative models. Another challenge pertains to the detailed information and textures generated by generative models. While they offer rich information, their strong prior distribution constrains the diversity of generated samples, primarily influenced by the choice of the training dataset. Hence, selecting an appropriate GAN for inverse problems using GAN as priors is crucial. Further investigation into transfer learning of pre-trained generative models can facilitate their adaptation to real-world data.

In our comparison of the four models, we have demonstrated that both SVGD and MALA stand out as the most efficient methods, as both methods use discrete points to represent distributions. However, there are opportunities to further enhance efficiency, particularly by reducing the number of particles or reduce the number of iterations. Increasing the number of particles and number of iterations can lead to a more accurate posterior distribution but comes with a trade-off of higher computational costs. Future studies could explore Gaussian mixture models, which parameterize the distribution using combinations of Gaussian distributions could provide a method to model the full posterior distribution more effectively. Additionally, machine learning-based PDE solvers, such as Physics-Informed Neural Networks (PINN) \citep{raissi2019physics, karniadakis2021physics} and Neural Operators \citep{li2020fourier}, can be considered as alternatives to traditional numerical PDE solving methods. These methods offer the potential to obtain the gradient of the likelihood function more efficiently, potentially improving the computational feasibility of high-dimensional inverse problem tasks.

Our study contributes to the theoretical understanding of seismic imaging by integrating deep generative model as prior for Bayesian inference methods. However, it is important to note that while our study focuses on theoretical advancements, implementing these techniques on real-world datasets presents its own set of challenges. Real datasets often introduce complexities and nuances that may not be fully captured in theoretical models. Additionally, in real-world applications, determining the appropriate noise level for the likelihood function is crucial for accurately modeling data uncertainties. While in our study, we use a fixed data noise level to provide realistic sampling from GAN, in practice, this noise level should be determined empirically from the data itself. Therefore, further research is warranted to explore the practical implementation of these theoretical advancements on real-world datasets. By combining the strengths and respective limitations of both deep generative models and Bayesian inference, we can gain new understanding of the complex processes in subsurface environments and help promote sustainable resource management.

\section{Conclusion}
To conduct stochastic seismic imaging and provide uncertainty regarding the inversion results using generative priors, we evaluate variational Bayesian inference methods, including inference network and normalizing flow and SVGD, and compare them with the conventional Bayesian inference, Markov chain Monte Carlo method MALA. The results indicate that SVGD and MALA, offers rapid solutions and can generate more variability in the posterior distribution, but provides limited number of posterior samples. In comparison, the neural network-based methods, inference network and normalizing flow, provide less variability but can provide continuous distributions as solutions and generate unlimited number of samples. Moreover, all the mentioned methods are able to represent the posterior distribution that contains the true model. Overall, we conclude that deep generative models can serve as prior distributions for different Bayesian inference methods and quantify uncertainty in full waveform inversion problem.

%Bibliography
% \bibliographystyle{unsrt}  
\bibliographystyle{abbrvnat}
\bibliography{references}

\end{document}